\documentclass[pdflatex,numbered]{manuscript}

\let\affil\relax
\usepackage{authblk}

\usepackage{graphicx}%
\usepackage{multirow}%
\usepackage{amsmath,amssymb,amsfonts}%
\usepackage{amsthm}%
\usepackage{mathrsfs}%
\usepackage[title]{appendix}%
\usepackage{xcolor}%
\usepackage{textcomp}%
\usepackage{manyfoot}%
\usepackage{booktabs}%
\usepackage{algorithm}%
\usepackage{algorithmicx}%
\usepackage{algpseudocode}%
\usepackage{listings}%

\usepackage{chngcntr}%
\usepackage{float}      
\usepackage{placeins}

\usepackage{soul}
\sethlcolor{yellow}

\theoremstyle{thmstyleone}%
\theoremstyle{thmstyletwo}%

\theoremstyle{thmstylethree}%

\usepackage{verbatim}

\makeatletter
\renewcommand\section{\@startsection{section}{1}{\z@}%
                                    {-16pt \@plus -6pt \@minus -3pt}%
                                    {8pt}%
                                    {\sectionfont}}
\renewcommand\subsection{\@startsection{subsection}{2}{\z@}%
                                       {-10pt \@plus -5pt \@minus -3pt}%
                                       {5pt}%
                                       {\subsectionfont}}
\renewcommand\subsubsection{\@startsection{subsubsection}{3}{\z@}%
                                          {-6pt \@plus -5pt \@minus -3pt}%
                                          {3pt}%
                                          {\subsubsectionfont}}
\makeatother
\begin{document}


\title[Article Title]{Unveiling healthcare-access inequality in Ghana using a multiscale approach.}

\author[1,*]{Miao Zeng}
\author[2,1]{Roberto Murcio}
\author[3]{Camilo Vargas-Ruiz}
\author[1]{Elsa Arcaute}
\affil[1]{The Bartlett Centre for Advanced Spatial Analysis (CASA), University College London, London, United Kingdom}
\affil[2]{School of Social Sciences, Birkbeck, University of London, London, United Kingdom}
\affil[3]{Malaria Atlas Project, The Kids Research Institute Australia, Nedlands, Australia}
\affil[*]{Corresponding author: ucbqmz7@ucl.ac.uk}

\abstract{
Achieving universal health coverage, as set out in Sustainable Development Goal 3.8, requires closing persistent geographic and socioeconomic gaps in healthcare access, especially in under-resourced settings across Africa. Healthcare-access inequality is shaped not only by poor local access, but also by limited connectivity to city- and regional-level services and opportunities. Conventional accessibility analysis can identify where poor access occurs, but not whether poorly served places form structurally disconnected pockets across scales. This paper therefore builds on and extends the percolation divergence tree framework to develop a connectivity-based multiscale approach for examining healthcare-access inequality in Ghana. It combines street-level accessibility mapping with the hierarchical structure of the road network to identify the scales at which inequality coincides with connectivity breaks. The results first show substantial inequality: around one quarter of the population lives more than 5 km from the nearest healthcare facility. Multiscale analysis further reveals distinct structural forms of poor access. In relatively well-connected, monocentric regions, local poor-access pockets emerge around metropolitan fringes despite overall regional advantage. In less well-connected, polycentric regions, poor access extends across larger subsystems, with local pockets nested within broader poorly served areas. These findings show that healthcare-access inequality reflects both local conditions and the hierarchical connectivity of the wider spatial system, which may also constrain marginalised communities' access to other key resources and services. The framework can inform targeted local interventions and policy coordination across local and regional scales.
}



\renewcommand{\maketitle}{%
  \vspace*{10mm}
  \begin{center}
  {\LARGE\bfseries Unveiling healthcare-access inequality in Ghana using a multiscale approach.\par}
  \medskip

  \large
  Miao Zeng\textsuperscript{1,\Large*},\quad
  Roberto Murcio\textsuperscript{2,1},\quad
  Camilo Vargas-Ruiz\textsuperscript{3},\quad
  Elsa Arcaute\textsuperscript{1}\par
  \smallskip

  \normalsize
  \textsuperscript{1,*}The Bartlett Centre for Advanced Spatial Analysis (CASA), University College London, London, United Kingdom.\par
  \textsuperscript{2}School of Social Sciences, Birkbeck, University of London, London, United Kingdom.\par
  \textsuperscript{3}Malaria Atlas Project, The Kids Research Institute Australia, Nedlands, Australia.\par
  \smallskip

  \small
  *Corresponding author(s). E-mail(s): \texttt{ucbqmz7@ucl.ac.uk};\par
  Contributing authors: \texttt{r.murcio@bbk.ac.uk}; \texttt{camilo.vargas@thekids.org.au}; \texttt{e.arcaute@ucl.ac.uk};\par
  \normalsize
  \end{center}
  \smallskip

  {\printabstract}
  {\printkeywords}
}

\maketitle


\section{Introduction}\label{sec1}

\noindent

Sustainable Development Goal (SDG) 3.8 calls for universal health coverage (UHC) \citep{WHOGHOSDG38}, yet progress in overall national coverage can mask persistent socioeconomic and geographic inequalities in healthcare access. Closing these gaps therefore remains a key challenge to achieving UHC \citep{Asamoah2026}. This challenge is particularly important in under-resourced settings across Africa, where infrastructure scarcity and long travel times can delay the diagnosis and treatment of diseases such as malaria \citep{Falchetta2020}. Although expanded prevention and control programmes have substantially reduced infection and mortality, substantial risks persist in poorly served communities with greater exposure and limited access to timely diagnosis, treatment, and prevention \citep{Bhatt2015}.

In this context, identifying poor-access pockets to provide targeted support is essential. Understanding why pockets of poor-access persist and remain difficult to address, therefore, requires situating them within a broader structure of spatial inequality, which also shapes other aspects of wellbeing, social mobility, and economic opportunity \citep{Deaton2001, Chetty2026, Sarkar2024}. Inequality is not merely the sum of individual disadvantages; it operates through neighbourhood and place-based effects that shape the life chances of whole communities beyond individual circumstances \citep{Chetty2026}. These effects also have a relational dimension: communities differ not only in their internal characteristics, but also in how they are connected to wider social and economic networks \citep{Chetty2022SocialCapital}. In spatial systems, such place-based inequality can be accumulated and reinforced over time  \citep{Sarkar2024}. Urbanisation and regional agglomeration are often part of this reinforcement process, through concentrating investment and opportunity in some places while deepening spatial exclusion elsewhere \citep{Sarkar2024}. Spatial connectivity provides a material counterpart to this relational view: relative connectivity breaks in transport-network can reveal how better-developed regions and less-developed regions are separated within the spatial system \citep{Arcaute2016}. Without a clear diagnosis of structural inequalities, policy interventions may reproduce or deepen the neglect and disconnection of poorly served communities. Over time, such polarisation between the well-connected “centre” and the poorly connected “periphery” can generate wider social and political backlash: turning the neglect of ``places that don't matter'' into a geography of revolt \citep{RodriguezPose2018}.


When it comes to healthcare-access inequalities, spatial accessibility is a commonly used indicator, especially where survey data are sparse or unevenly distributed. Access to essential infrastructure, amenities, and services is itself spatially uneven, so accessibility analysis helps reveal where access is good or poor. Globally, accessibility measures capture broad disparities between the Global North and South; for example, the Global South has far lower per capita infrastructure provision, and walking times to services in sub-Saharan Africa remain high \citep{Wu2025}. At regional scales, travel-time mapping has highlighted persistent rural--urban gaps in access to critical infrastructure \citep{Weiss2018}, while recent work shows that contemporary urbanisation can widen rather than close between-region infrastructure gaps \citep{Pandey2025}. Within cities, high-resolution accessibility mapping has revealed intra-urban gradients and pockets of exclusion, including access gaps associated with health disparities and deviations from the ``15-minute city'' ideal \citep{Tu2025, Bruno2024, Zhou2026}. However, most accessibility analysis remains single-scale or locally defined. It can identify locations or regions with poor access, but it cannot directly show whether poorly served communities form spatially concentrated pockets or at what scale such exclusion emerges.

This limitation reflects deeper methodological limitations in conventional accessibility analysis. Most accessibility metrics are fundamentally local: they pair each location with nearby services and treat locations as independent data points \citep{LuoWang2003, Wang2012, Tao2020}. To identify group-level deprivation or inequality, these local measurements are usually aggregated at the level of administrative units or statistical averages \citep{Openshaw1984, Attili2020}. This can obscure the structural pattern of inequality in two ways. The first is the aggregation problem: the use of exogenous and arbitrarily-defined aggregation units, such as administrative areas, can lead to the Modifiable Areal Unit Problem \citep{Openshaw1984, Marin2024}. This problem is especially important for healthcare access, because facility catchments often cut across formal administrative boundaries \citep{Tao2020, Jia2017}. The second is the representation problem: group-level means can oversimplify inequality by hiding distributional variance \citep{Dagum1997, Attili2020, Silber1989}. They can hide the spatial concentration of poorly served communities that administrative boundaries fail to capture \citep{ReySmith2013}. As a result, pockets of poor access may remain invisible within apparently well-served districts or regions \citep{Openshaw1984, Pradhan2003}.Therefore, a more realistic diagnosis needs to move beyond single-scale proximity mapping and administrative aggregation. Some work has advocated for a multiscale accessibility profile \citep{KaplanOmer2022}, while others have argued that the key question is not only where poor access is located, but at what scale inequality emerges \citep{DuranNebreda2025}.


This methodological concern becomes especially important in Africa, where healthcare-access inequality is embedded within a wider structure of spatial hierarchy and infrastructure scarcity. Traditional accessibility analysis may help identify broad areas of poor access, but it can still miss how poorly served communities are concentrated, separated, and embedded within wider uneven spatial systems. Africa's settlement systems are often bottom-heavy, with most settlements small or intermediate in size and only a few metropolitan centres dominating the upper end of the hierarchy \citep{Heinrigs2020}. Yet even these metropolitan centres often expand through low-density and inefficient growth, without generating the agglomeration benefits expected at their scale \citep{Xu2025}. This uneven hierarchy is matched by severe infrastructure inequality: healthcare resources are scarce and spatially concentrated, while road connectivity remains limited and fragmented. More than 170 million people live more than two hours from the nearest hospital \citep{Falchetta2020}, and sparse road networks create connectivity breaks between urban cores and peripheral settlements, as well as between more-developed urban agglomerations and wider rural regions \citep{Bettencourt2025, PrietoCuriel2022, Heinrigs2020}.

In this work, we took Ghana as case of study for examining multiscale healthcare-access inequality within the African context. Its human settlements and service systems show a similar spatial hierarchy: population, infrastructure, and public services are concentrated around a few major urban centres, including Accra, Kumasi, and Tamale, while many smaller towns and rural settlements are distributed across lower-density areas with weaker road connectivity and more limited service capacity.


The central question of this paper is therefore not only \textit{where are poor-access locations?}, but \textit{at what scale does structural healthcare-access inequality emerge?} More specifically, we ask whether poorly served locations form poor-access pockets that are separated from wider and better-served systems. Answering this question requires a framework that can identify both the spatial distribution of poor access and the connectivity structure within which poor-access populations are embedded.

Hence, we extend the Percolation Divergence Tree framework\citep{Neira2025} into a connectivity-based multiscale diagnostic framework for identifying structural healthcare-access inequality. This paper extends the Percolation Divergence Tree framework from categorical population compositions to distributions of continuous spatial variables, enabling connectivity-based multiscale analysis beyond segregation to broader forms of spatial heterogeneity and inequality. in  The framework starts with measuring street-level healthcare accessibility along the road network and combines it with population distribution, allowing poor access to be measured at high spatial resolution, including in remote and lower-density areas where coarse surveys or administrative averages can miss local conditions. Drawing on hierarchical percolation theory, we then identify connectivity-defined communities endogenously and record how separate components merge into larger connected systems across different scales \citep{Arcaute2016, Rozenfeld2011}. This bottom-up approach avoids relying on arbitrary administrative units and instead uncovers spatial clusters across a continuum of connectivity scales \citep{Arcaute2016, Marin2024}. To characterise inequality within this hierarchy, we move beyond group averages and compare accessibility distributions between connectivity-defined communities across multiple scales \citep{Murcio2015, Murcio2025}. Specifically, we apply the Generalised Jensen--Shannon Divergence (GJSD) to the hierarchical connectivity tree, following the framework developed by \citep{Neira2025}, which allows us to identify the scales where healthcare-access inequality is more prominent. By measuring distributional divergence on the percolation tree \citep{Neira2025, Murcio2015}, the framework identifies meaningful scales where healthcare-access inequality coincides with road-network connectivity break \citep{Neira2025, DuranNebreda2025}. Applied to Ghana, this framework identifies where poorly served populations are located, how many people are affected, where poor-access pockets or regional subsystems coincide with connectivity breaks, and whether these inequalities appear as local, regional, or nested patterns. In doing so, it has the potential to inform policy discussions on targeted local intervention and multi-scale coordination in settings where infrastructure scarcity and hierarchical connectivity structure jointly shape healthcare access.

\section{Methods}\label{sec2}

\subsection{Study area \& data}\label{subsec:study_area_data}

Ghana is used as the study area because its settlement and infrastructure systems are spatially uneven and hierarchically organised. Service infrastructure is concentrated around major urban centres, such as Greater Accra, Kumasi, and Tamale, while much of the population is dispersed across smaller towns and rural settlements in lower-density areas. This spatial contrast makes Ghana a suitable case for examining how healthcare access varies across both local and regional connectivity contexts.

Alongside this settlement hierarchy, Ghana's health system is organised as a tiered referral system, with Community-based Health Planning and Services (CHPS) forming an important community-level entry point to primary care, followed by health centres and hospitals at district, regional, and teaching levels. CHPS services are delivered by Community Health Officers (CHOs), trained nurses or midwives who provide primary care and referral services both at CHPS compounds and through outreach visits to households and communities \citep{Asiedu2023}. This outreach component extends healthcare provision beyond fixed facilities, particularly in rural and poorly served areas. Because our analysis measures accessibility from the geographic locations of healthcare facilities, it includes access to fixed CHPS compounds but not the wider geographic reach of CHPS outreach services. The resulting measure should therefore be interpreted as facility-based rather than complete healthcare access.

We use four main spatial datasets. First, road data were obtained from OpenStreetMap and converted into a road-network graph, in which road intersections are represented as nodes and road segments connecting them are represented as edges. Second, healthcare facility locations are obtained from the Ministry of Health and complemented with facility data from the Malaria Atlas Project. After cleaning, the dataset retains health facilities across different levels of the healthcare system, including clinics, health centres, CHPS compounds, reproductive and child health (RCH) facilities, polyclinics, and hospitals at district, regional, municipal, metropolitan, and teaching levels, while administrative and other non-clinical facilities were excluded (Figure~\ref{Appfig_Healtcare_Facilities}).  Third, population distribution is represented using the WorldPop constrained gridded population dataset \citep{Tatem2017WorldPop, WorldPopR2024B} (2023 population estimates, 2024B release), which provides population counts at approximately 100 m spatial resolution (Figure~\ref{Appfig:worldpop_map}). Fourth, administrative boundaries of the sixteen regions in Ghana, are used for mapping and contextual interpretation, but not as the primary units of analysis.

All spatial datasets are cleaned and projected into a common coordinate system. Healthcare facilities are linked to their nearest road intersections, and the WorldPop grid is spatially joined with the street-intersection network. Population counts from grid cells are then assigned to their nearest intersections, producing a population-weighted set of road-network locations. These processed datasets support the three analytical steps described below: street-level healthcare accessibility, road-network percolation, and the identification of accessibility inequality on the percolation tree.

\subsection{Street-level healthcare accessibility}\label{subsec:flat_access}

The first step is to measure and map spatial inequalities in healthcare access at a high spatial resolution. We therefore measure access from each street intersection rather than from administrative units. This street-level measure allows the analysis to show where well-served and poorly served locations are distributed, and, after population weighting, how many residents live within different distance bands of healthcare access.

To quantify healthcare accessibility, we calculate the shortest walking distance from each street intersection to the nearest healthcare facility along the road network. This choice provides a consistent national baseline, because complete and standardised data on public-transport routes, schedules, fares and service reliability are not available for all locations in Ghana. More importantly, walking distance does not assume that residents have access to private vehicles or reliable public transport. It therefore follows an equity-oriented perspective, focusing on the physical proximity to healthcare that would matter most for residents with limited transport options.

For each street intersection $i$, distance-based accessibility is measured as the shortest network distance to the nearest healthcare facility, computed using Dijkstra's algorithm:
\begin{equation}
A(i) = \min_{f \in \mathcal{F}} d_G(i, f)
\label{eq:accessibility}
\end{equation}

\noindent where $\mathcal{F}$ is the set of healthcare facilities, $d_G(i, f)$ denotes the walking distance from intersection $i$ to facility $f$ along the road network $G$. Hence, $A(i)$ is the minimum of these distances, that is, the walking distance to the nearest facility. We use 5~km as a reference threshold for poor healthcare access. In the WHO UHC monitoring report, the share of the population living within 5~km of a health facility is used as a spatial-access measure associated with the coverage of primary health services \citep{WHO2019PHCUHC}. This distance is also close to a one-hour walking distance. We therefore use $A(i) > 5~\mathrm{km}$ to identify intersections with poor spatial access to healthcare.

We then spatially overlay the WorldPop 2024 population grid with the road network (Figure~\ref{Appfig:worldpop_map}). The population count from each grid cell is assigned to the nearest street intersections, giving each intersection a population weight $w(i)$. This produces a population-weighted accessibility distribution, where each intersection has both an accessibility value $A(i)$ and a population weight $w(i)$. The national accessibility distribution therefore reflects not only how far each street intersection is from healthcare, but also how many people live near that location.
\subsection{Percolation on road network: hierarchical connectivity structure}\label{subsec:perco}

The street-level accessibility measure describes where access is good or poor, but it treats each location as an isolated point. It does not show whether poorly served locations form connected pockets, or how these pockets are separated from wider and better-served systems. To recover this spatial context, we then represent the settlement system as a connected road-network rather than as a set of independent locations.

This road-network connectivity structure is not uniform across territorial space. We therefore use distance-threshold percolation to reveal how this connectivity structure is both heterogeneous and hierarchical: dense urban cores first emerge as connected components at small thresholds, then grow into sub-national systems at intermediate thresholds, and eventually merge into a national whole.

Let the road network be represented as a graph $G=(V,E)$, where $V$ is the set of street intersections and $E$ is the set of road segments. Each road segment $e \in E$ has a length $\ell(e)$. At a distance threshold $\tau$, we define the active subgraph as:

\begin{equation}
G_{\tau} = (V, E_{\tau}), \quad E_{\tau} = \{e \in E \mid \ell(e) \leq \tau\}.
\label{eq:active_graph}
\end{equation}

At each threshold $\tau$, the active subgraph decomposes into connected components, or clusters:

\begin{equation}
S(\tau) = \{C_1^{\tau}, C_2^{\tau}, \ldots, C_{k_{\tau}}^{\tau}\},
\label{eq:clusters}
\end{equation}

\noindent where $S(\tau)$ is the set of all clusters at threshold $\tau$, $C_j^{\tau}$ is the $j$-th connected component, and $k_{\tau}$ is the total number of clusters. We start with $\tau_0 = 100\,\mathrm{m}$ and increase the threshold in fixed steps of $\Delta\tau = 100\,\mathrm{m}$. At each step, we compute the connected components of $G_{\tau}$. The process continues until the largest connected component contains more than 95\% of all intersections, indicating near-complete national integration.

\begin{figure}[htbp]
  \centering
  \includegraphics[width=0.95\textwidth]{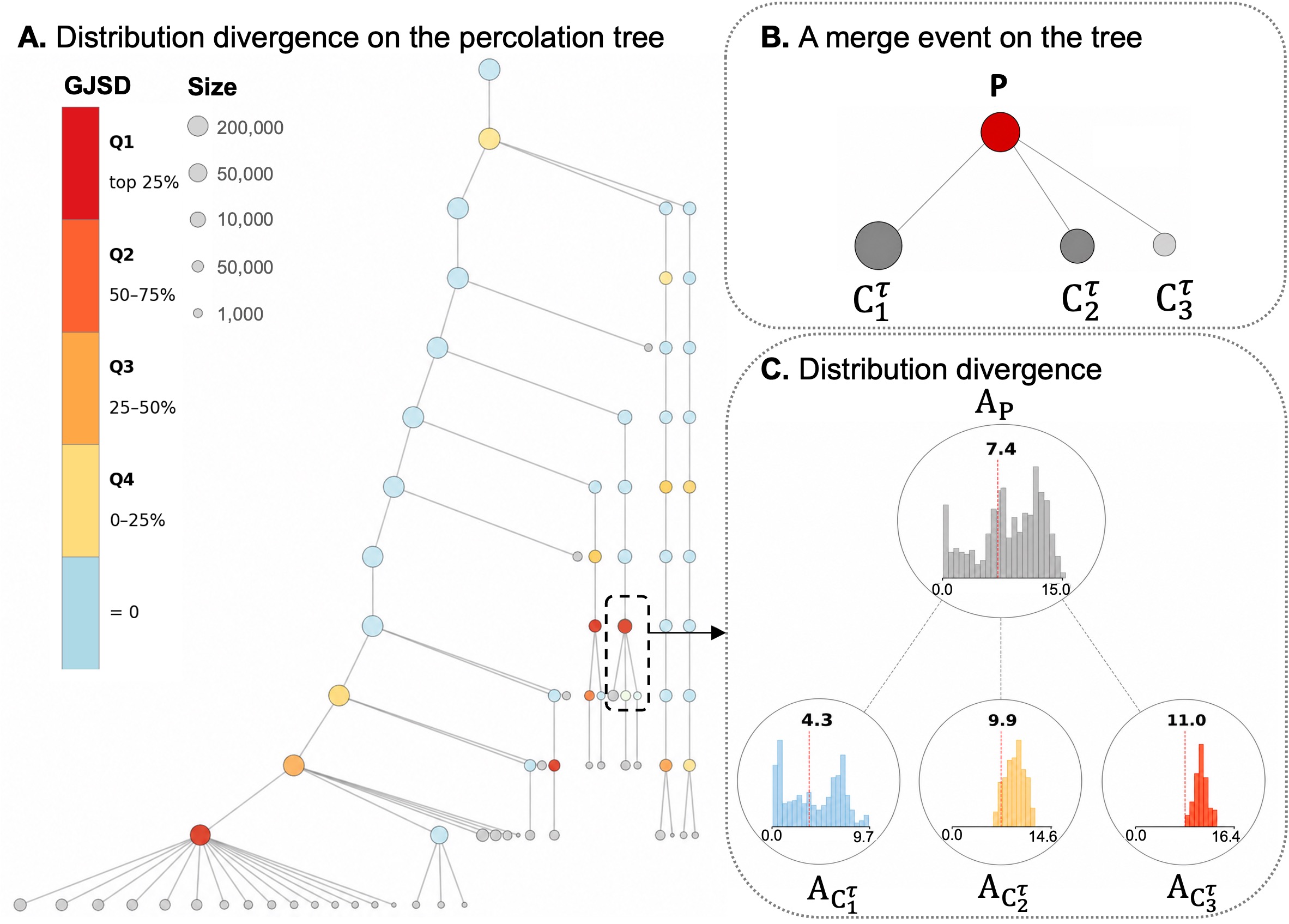}
  \caption{\textbf{Measuring accessibility-distribution divergence on the percolation tree.}
  \textbf{(A)} Example percolation tree, where node size represents cluster size and node colour represents the GJSD value at each merge event.
  \textbf{(B)} Schematic merge event: child clusters $C_1^{\tau}$, $C_2^{\tau}$, and $C_3^{\tau}$ merge into a parent cluster $P$ at a higher distance threshold.
  \textbf{(C)} Accessibility distributions for a representative high-GJSD merge event. The grey histogram shows the parent distribution $A_P$, and the coloured histograms show the child-cluster distributions. High GJSD indicates that the child clusters have clearly different healthcare-access profiles.}
  \label{fig:method_perco_gjsd}
\end{figure}

The successive cluster sets $S(\tau)$ form a hierarchical percolation tree (Figure~\ref{fig:method_perco_gjsd}A). Each node in the tree represents a connected spatial cluster identified at a given distance threshold. At small thresholds, the road network is fragmented into many small local clusters. As $\tau$ increases, nearby clusters merge into larger connected systems. Each parent--child relationship on the tree therefore records a merge event, showing how smaller clusters become part of a larger connected cluster at a higher threshold (Figure~\ref{fig:method_perco_gjsd}B).

The tree also captures how uneven this connectivity structure is. Node size represents cluster size, and the relative sizes of sibling branches indicate whether merging subsystems are similar or very different in scale. Strong size asymmetry between branches suggests a dominant subsystem absorbing much smaller surrounding systems, whereas more balanced branches suggest multiple subsystems of comparable scale. This structural variation provides an important context for spatial inequality, as uneven connectivity may be associated with uneven service provision and access to infrastructure. The percolation tree therefore not only describes how connectivity is organised, but also provides the structural basis for comparing healthcare accessibility across communities and scales.

\subsection{Identifying health-care access inequality on the percolation tree}\label{subsec:divergence_on_tree}

Here, we adapt the Percolation Divergence Tree framework \citep{Neira2025} from comparing categorical population compositions to comparing empirical distributions of a continuous-valued spatial attribute—healthcare accessibility—across the connectivity hierarchy. We then use this hierarchy to identify where healthcare-access inequality emerges. Not every connectivity break is an inequality break: some structurally separated clusters have similar healthcare-access profiles, while others differ sharply because healthcare facilities are unevenly distributed across the split. We therefore compare accessibility profiles across the percolation tree to identify merge events where structural separation and healthcare-access inequality occur together.

For each cluster $C_j^{\tau}$, we aggregate the street-level accessibility values of all intersections in that cluster. This creates an intersection-level accessibility distribution for each cluster:
\begin{equation}
\mathcal{A}_{C_j^{\tau}} = \{A(i) \mid i \in C_j^{\tau}\}.
\label{eq:cluster_accessibility}
\end{equation}

\noindent where $\mathcal{A}_{C_j^{\tau}}$ is the set of accessibility values for all intersections in cluster $C_j^{\tau}$, and $A(i)$ is the shortest walking distance from intersection $i$ to the nearest healthcare facility. This aggregation allows each connectivity-defined cluster to be described by the distribution of healthcare access within it.  At each merge event, a parent cluster $P$ is formed from a set of child clusters $C_1^{\tau}, C_2^{\tau}, \ldots, C_n^{\tau}$ identified at the previous threshold step. We denote the accessibility distribution of the parent cluster as $\mathcal{A}_{P}$, and the accessibility distribution of each child cluster as $\mathcal{A}_{C_i^{\tau}}$ (Figure~\ref{fig:method_perco_gjsd}C).

To compare accessibility distributions across the hierarchy, we use the Generalised Jensen--Shannon Divergence (GJSD). To represent the continuous accessibility variable as a probability distribution, accessibility values are discretised using a common set of bins across all clusters. For each cluster, the proportion of intersections falling within each bin defines its empirical accessibility distribution. GJSD measures how different the child accessibility distributions are from the merged parent distribution:

\begin{equation}
\mathrm{GJSD}(P) = H(\mathcal{A}_{P}) - \sum_{i=1}^{n} \pi_i H(\mathcal{A}_{C_i^{\tau}}),
\label{eq:gjsd}
\end{equation}

\noindent where $H(\mathcal{A}_{P})$ is the Shannon entropy of the parent cluster's accessibility distribution, $H(\mathcal{A}_{C_i^{\tau}})$ is the entropy of the $i$-th child cluster's accessibility distribution, and $\pi_i$ is the relative size of child cluster $C_i^{\tau}$ measured by the number of intersections:

\begin{equation}
\pi_i = \frac{|C_i^{\tau}|}{\sum_{j=1}^{n} |C_j^{\tau}|}.
\label{eq:gjsd_weight}
\end{equation}

Merge events with GJSD values equal to, or close to, zero are shown as light-blue nodes in Figure~\ref{fig:method_perco_gjsd}A. These cases indicate structural disconnection without a clear difference in healthcare-access profiles. Non-zero GJSD values are coloured by quantile, with Q1 representing the highest GJSD quantile. We focus on Q1 nodes, shown in the darkest red, because they mark merge events in which structural separation and healthcare-access inequality occur together. Figure~\ref{fig:method_perco_gjsd}C shows a high-GJSD merge event, in which child clusters have clearly different healthcare-access profiles: some are well served overall, while others are less well served.  We then select Q1 merge events from these to identify meaningful poor-access pockets. Specifically, we retain only those in which at least one worse-served child cluster has more than 1,000 people with $A(i) >5~\mathrm{km}$. This criterion ensures that the selected scales show high distributional divergence and at least one substantial poorly served cluster.

The position of flagged nodes in the tree indicates the scale of the poor-access problem. Lower-level flagged nodes identify local poor-access pockets, where a small cluster differs from a nearby better-served system. Higher-level flagged nodes identify broader regional poor-access subsystems, where a larger connected subsystem remains separated from the wider system until a higher threshold. Repeated flagged nodes along the same branch indicate nested deprivation, where local poor-access pockets are embedded within a wider poorly served regional system. Together, this multi-scale reading distinguishes whether poor access is local, regional, or nested across scales, showing why it cannot be understood, or addressed, through a single uniform spatial scale.
\section{Results}\label{sec3}

\subsection{Street-level inequality in healthcare access}\label{subsec:res_accs}

\FloatBarrier

\begin{figure}[H]
  \centering
  \includegraphics[width=0.7 \textwidth]{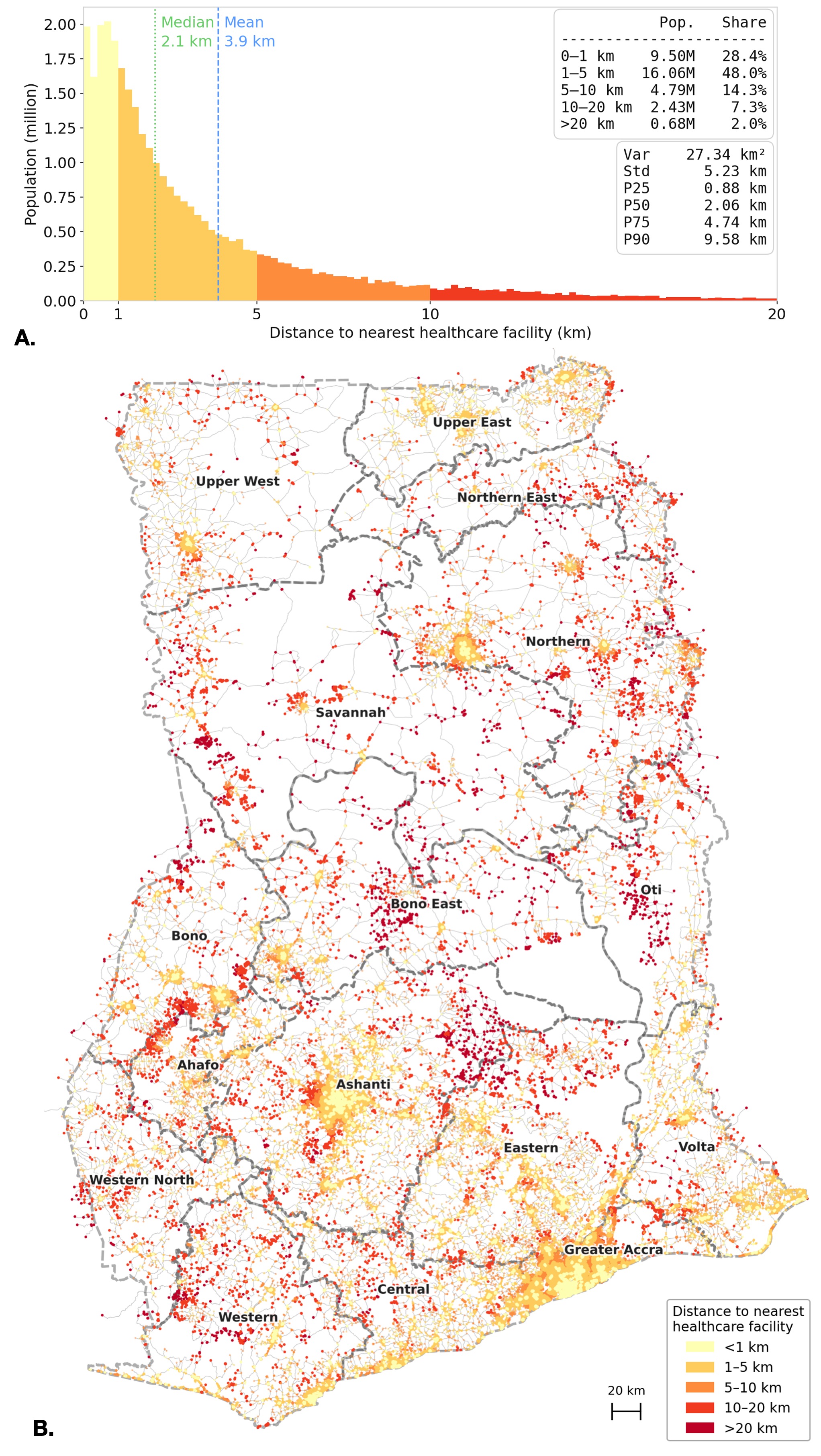} 
  \caption{\textbf{National healthcare accessibility in Ghana's.}
  \textbf{(A)} Walking distance(km) from each street intersections to the nearest healthcare facility in Ghana.
  \textbf{(B)} Distribution of population by distance to nearest healthcare facility.}
  \label{fig:accessibility}
\end{figure}

We first present the access to healthcare at the street level, which gives the most local view of accessibility before any spatial aggregation is applied. Figure~\ref{fig:accessibility} shows street-level inequality in healthcare access in Ghana, combining the spatial map and the population-weighted distance distribution. Using the 5~km reference threshold defined in the Methods, Ghana appears relatively well served at first glance: the national mean distance is below 5~km. However, this average gives an overly positive picture and masks substantial internal inequality. Figure~\ref{fig:accessibility}B shows a sharp contrast between the well-served end of the distribution and the underserved tail. At the well-served end, about 28.4\% of the population lives within 1~km of a healthcare facility. These residents are mainly concentrated in towns and urban centres, including Greater Accra, Kumasi, Tamale, and other regional and local centres. In the underserved tail, access is much worse. Around 25\% of the total population, or more than 8 million residents, live more than 5~km from the nearest facility. Nearly 10\%, or more than 3 million residents, live more than 10~km away.

The map in Figure~\ref{fig:accessibility}A adds the spatial context behind this poorly served tail. It shows where poorly served populations are located and that they are not randomly distributed. Some poor-access areas appear as small but high-density concentrations, while others are more dispersed and extend across larger peripheral areas. This suggests that the long-distance tail contains different kinds of access problems. However, the flat map still treats each location as an isolated point. It cannot identify distinct poor-access pockets in different spatial contexts, yet this distinction is important for understanding how poor-access patterns are formed and for designing more targeted interventions.

\FloatBarrier
\subsection{Hierarchical connectivity and multi-scale inequality in healthcare access }\label{subsec:res_hier}

\begin{figure}[p]
    \centering
    \includegraphics[width=1.0\textwidth]{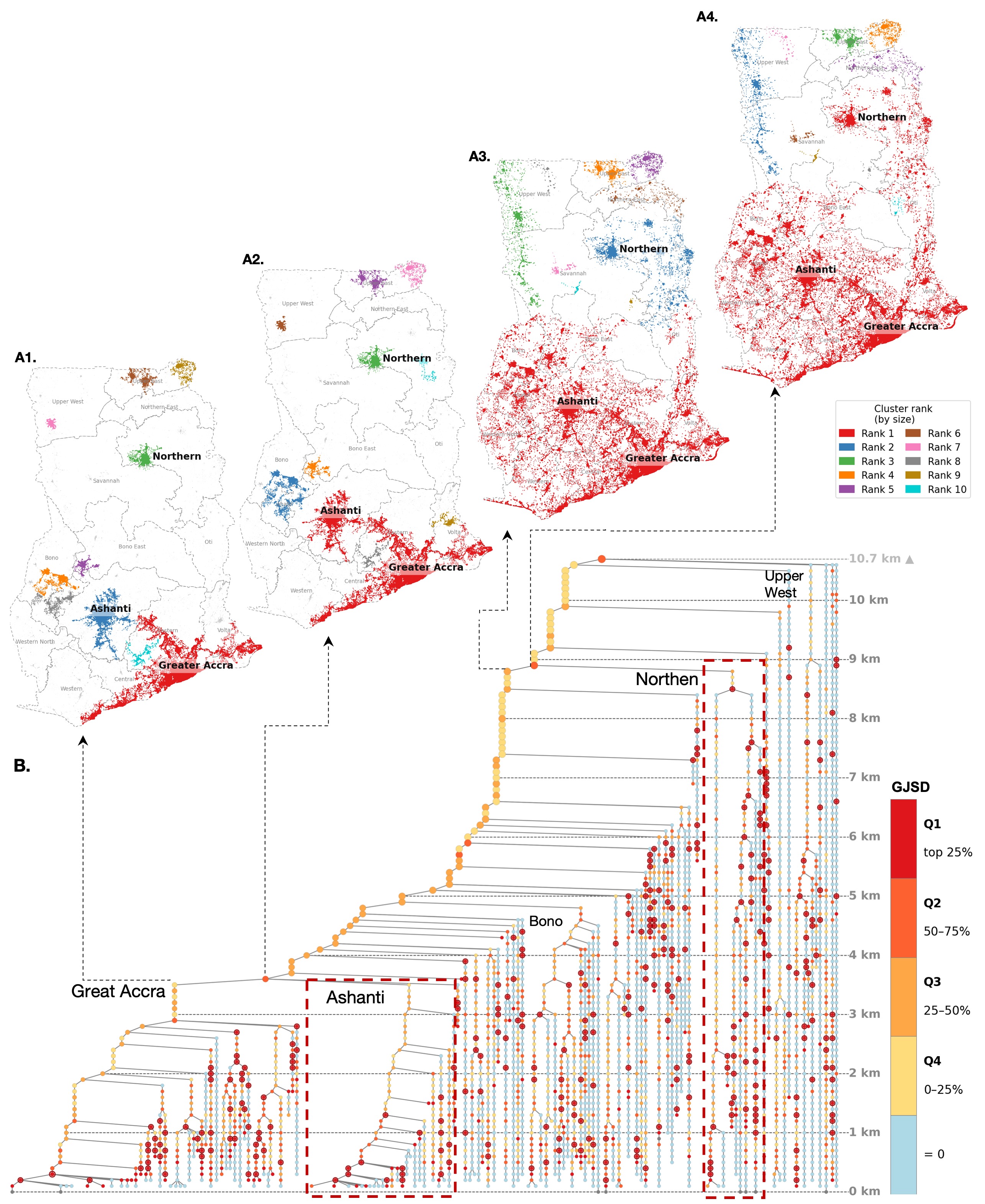}
    \caption{\textbf{Hierarchical connectivity and multi-scale healthcare-access inequality in Ghana.}
    \textbf{(A)} Spatial configuration of the ten largest clusters at four selected percolation thresholds. Cluster colours in Panel~A indicate size rank in descending order, with Rank~1 representing the largest cluster at each threshold. From left to right, Panels A1--A4 show the cluster structure at 3.5, 3.6, 8.8, and 8.9~km, respectively. Panels A1--A2 show the Ashanti transition, where the Ashanti system joins the national backbone at a lower threshold. Panels A3--A4 show the Northern transition, where the Northern system joins the national backbone at a higher threshold.
    \textbf{(B)} Hierarchical percolation tree of Ghana's road network. Each node represents a connectivity-defined cluster identified at a given percolation threshold; node size is proportional to cluster size. Node colour encodes the GJSD value at each merge event, discretised into five quantile classes from Q1 to Q5, with Q1 representing the highest GJSD quantile; light-blue nodes have GJSD $=0$. Circled dark-red nodes mark flagged meaningful scales, where high distributional divergence coincides with a sufficiently large poor-access pocket.}
    \label{fig:GJSD_Global}
\end{figure}

We next move from the flat accessibility map to the hierarchical connectivity structure of Ghana's road network. Figure~\ref{fig:GJSD_Global}B represents this hierarchical and heterogeneous connectivity structure in Ghana. Greater Accra appears early as the primary core of the largest connected component (LCC), and the LCC then expands along the southern backbone. Ashanti joins this backbone at one of the major lower-level transitions ($\sim$3.5~km), while Bono and Bono East join at intermediate levels ($\sim$4~km). The Northern system joins only at a higher level ($\sim$9~km), and Upper West and Upper East remain separate until later stages ($>$10~km). These transitions show that different regions are integrated into the national road-network system at different connectivity scales.

The maps above the tree (Figure~\ref{fig:GJSD_Global}A) illustrate the spatial meaning of two key transitions that are used as examples later in the case studies. Each pair of maps shows the cluster structure immediately before and after a regional subsystem joins the national backbone. The Ashanti transition (Figure~\ref{fig:GJSD_Global}A1--A2) represents a lower-level transition: a dense and compact regional system joins the national backbone at a relatively short threshold. Meanwhile, the Northern transition (Figure~\ref{fig:GJSD_Global}A3--A4) represents a higher-level transition: a more remote and dispersed regional system joins the national network only at a larger threshold. These examples show that regional systems enter the national connectivity structure at different thresholds and with different spatial forms.


Besides topology, we then examine which connectivity breaks also correspond to healthcare-access inequality. In Ghana, most merge events are light-blue nodes with GJSD $=0$ (Figure~\ref{fig:GJSD_Global}B). This means that many structural separations do not produce clear accessibility divergence. The flagged nodes therefore identify the smaller set of meaningful scales where high distributional divergence coincides with a substantial poor-access pocket.

The position of these flagged nodes on the tree reveals different forms of inequality. Lower-level flagged nodes indicate local poor-access pockets embedded within densely-connected systems. Higher-level flagged nodes indicate broader regional poor-access subsystems that remain separated from the national backbone until larger thresholds. These higher-level subsystems generally correspond to larger affected populations and wider territorial areas. Repeated flagged nodes along the same branch indicate nested inequality, where local poor-access pockets are embedded within wider poorly served regional systems. The next section illustrates these patterns through the Ashanti and Northern case studies.

\FloatBarrier
\subsection{Regional case studies: different forms of inequality across scales}\label{subsec:res_case}
To illustrate how the hierarchical diagnosis works in practice, we examine two contrasting regional systems: Ashanti and the Northern Region. Ashanti represents a relatively well-connected and monocentric system, while the Northern Region represents a more weakly connected and polycentric system. Comparing the two cases shows how healthcare-access inequality takes different forms across connectivity contexts and spatial scales.

\subsubsection{Ashanti: healthcare-access inequality in a well-connected region}\label{subsec2.1}
\begin{figure}[htbp]
  \centering
  \includegraphics[width=1.0\textwidth]{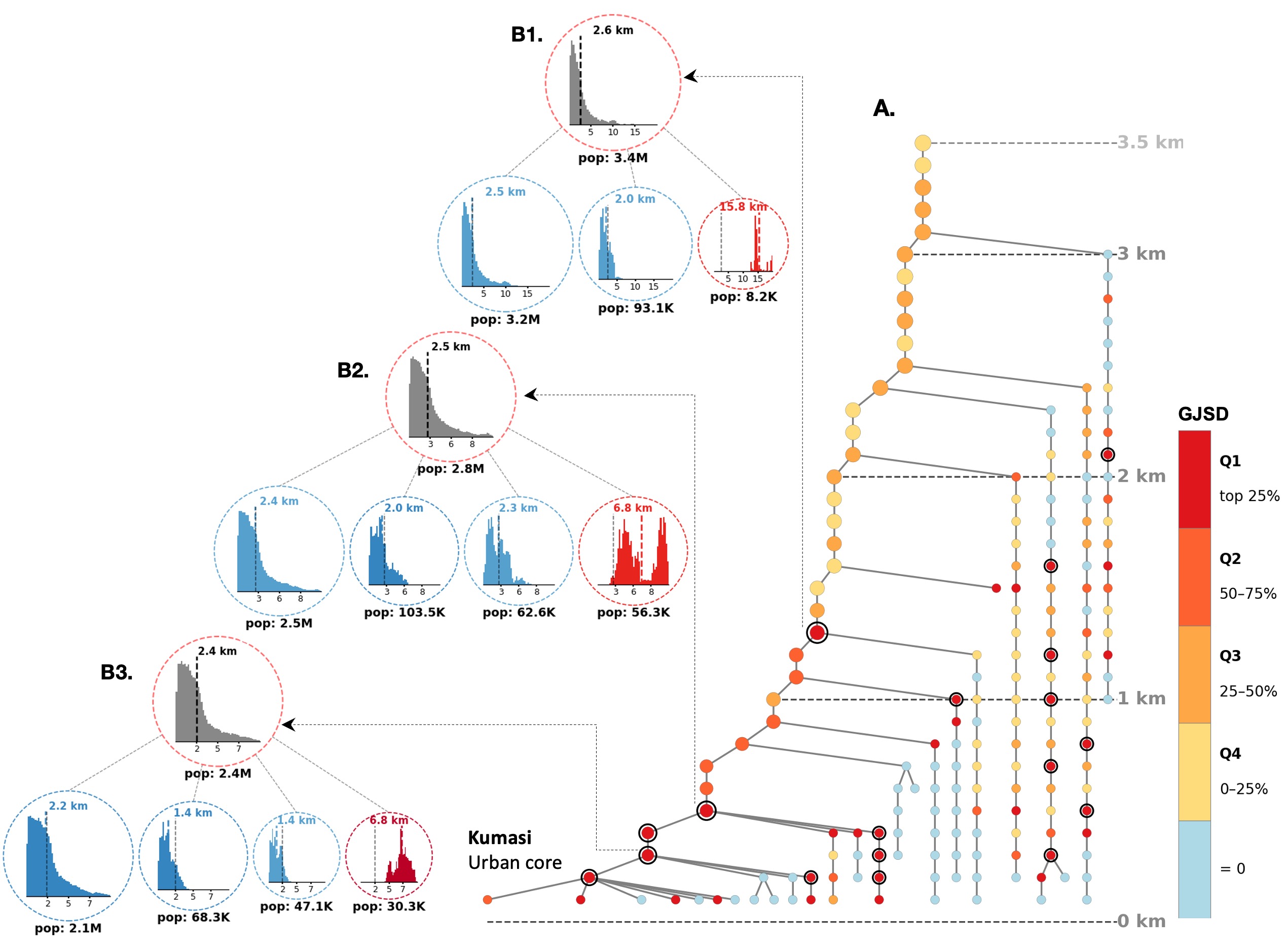}
  \caption{\textbf{Ashanti subtree and flagged scales of healthcare-access inequality.}
  \textbf{(A)} Percolation subtree for the Ashanti region, coloured by GJSD quantile as in Figure~\ref{fig:GJSD_Global}. Light-blue nodes have GJSD $=0$, and circled nodes mark flagged merge events with high internal inequality and substantial poor access pocket.
  \textbf{(B)} Healthcare-access distributions for three selected flagged merge events in the Ashanti subtree: B1, B2, and B3. Each panel shows the parent distribution in grey and the child-cluster distributions in colour. Child clusters are ordered from left to right by cluster size. Blue indicates better-served child clusters, while red/orange indicates worse-served child clusters. Dashed vertical lines mark mean distance to the nearest healthcare facility, and labels report mean distance and population size. Across the three examples, the largest and leftmost child cluster corresponds to the Kumasi urban core, while the smaller worse-served child clusters represent peripheral communities and settlements that merge with the core at different distance thresholds.}
  \label{fig:Subtree_1_Ashanti}
\end{figure}

Ashanti illustrates the first pattern: a monocentric and relatively well-connected region (Figure~\ref{fig:Subtree_1_Ashanti}). Ashanti joins the national backbone at a relatively short threshold of 3.5~km, indicating that the region is generally densely connected. Its subtree is compact and strongly centred on Kumasi, Ghana's second-largest city and the major metropolitan core of the region. In Figure~\ref{fig:Subtree_1_Ashanti}A, the main left-side backbone represents the growth of the Kumasi-centred system, while the smaller right-side branches represent surrounding towns and settlements. The persistent size asymmetry between this backbone and the side branches shows the monocentric structure of Ashanti: no other centre grows to a scale comparable to Kumasi.

Although Ashanti is well connected overall, the combined percolation-tree and GJSD analysis reveals clear internal inequality. The flagged nodes are mostly located at lower levels of the tree (Figure~\ref{fig:Subtree_1_Ashanti}B1--B3), where the Kumasi-centred core merges with smaller fringe clusters. These merge events show a clear accessibility split: the large Kumasi-centred core has substantially better healthcare access, while the smaller peripheral child clusters are systematically worse served. The Appendix maps (Appendix Figure~\ref{Appfig_Example_Ashanti}) show that these worse-served clusters are mainly located around the metropolitan fringe and nearby satellite settlements. This pattern is consistent with a mismatch between the expansion of surrounding settlements and the spatial distribution of healthcare facilities. The analysis therefore identifies local poor-access pockets that would otherwise be hidden by Ashanti's relatively advantaged regional average.
\FloatBarrier
\subsubsection{Northern Region: healthcare-access inequality in a weakly connected region}\label{subsec2.2}
\begin{figure}[htbp]
  \centering
  \includegraphics[width=1\textwidth]{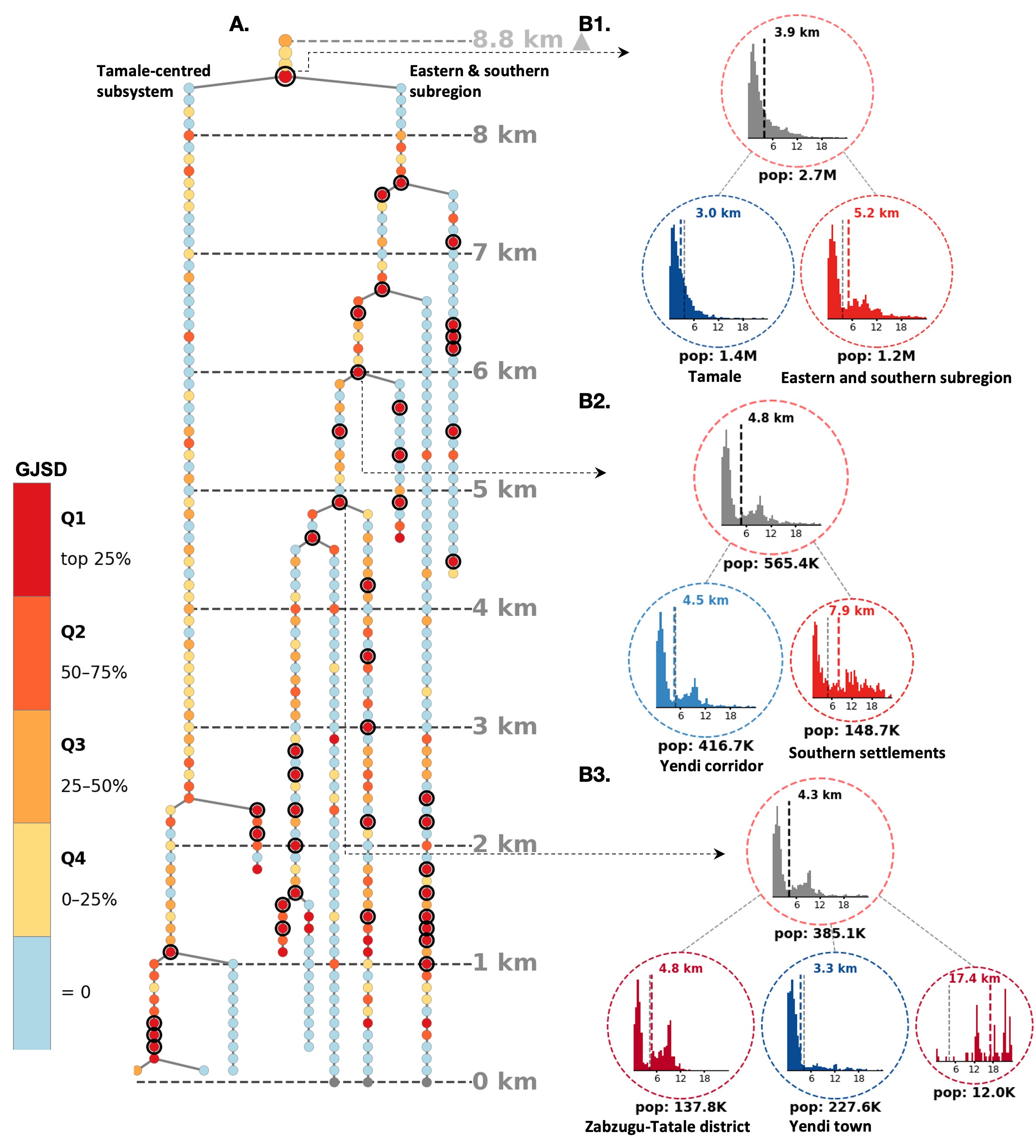}
  \caption{\textbf{Northern Region subtree and flagged scales of healthcare-access inequality.}
  \textbf{(A)} Percolation subtree for the Northern Region, coloured by GJSD quantile as in Figure~\ref{fig:GJSD_Global}. Light-blue nodes have GJSD $=0$, and circled nodes mark flagged merge events with high internal inequality and a substantial poor-access pocket.
  \textbf{(B)} Healthcare-access distributions for three selected flagged merge events: B1, B2, and B3. Grey indicates the parent distribution, blue indicates better-served child clusters, and red/orange indicates worse-served child clusters. Dashed vertical lines mark mean distance, and labels report population size. }
  \label{fig:Subtree_2_Northern}
\end{figure}

The Northern Region illustrates the second pattern: healthcare-access inequality in a weakly connected and polycentric region (Figure~\ref{fig:Subtree_2_Northern}). The Northern subtree joins the national backbone only at a much larger threshold of 8.8~km, showing that this regional system is relatively remote in the national system. Unlike Ashanti, the Northern region is not organised around one dominant metropolitan core. Several merge events in Figure~\ref{fig:Subtree_2_Northern}A show more balanced branch sizes, suggesting multiple sub-regional systems of comparable scale. The clearest example is the high-level merge event in Figure~\ref{fig:Subtree_2_Northern}B1, where two similarly sized sub-regional systems merge: the Tamale-centred subsystem and the more dispersed eastern and southern corridor system.

The GJSD pattern shows that inequality in the Northern region is not limited to local pockets. On the one hand, local poor-access pockets are still identified around major urban centres, such as Tamale, similar to the periphery settlements seen around Kumasi. On the other hand, flagged nodes also appear at much higher levels of the tree (Figure~\ref{fig:Subtree_2_Northern}B1--B3). For example (Figure~\ref{fig:Subtree_2_Northern}B1), regional disconnection is also related with regional level inequality. The better-served Tamale-centred subsystem has a mean distance of 3.0~km and a population of about 1.4 million, while the worse-served eastern and southern system (Appendix Figure~\ref{Appfig_Example_Northern}B1), has a mean distance of 5.2~km and a population of about 1.2 million. The two systems are therefore comparable in population size, but sharply different in healthcare accessibility.

Tracing the worse-served branch further down the Northern subtree reveals a nested pattern of poor-access inequality (Figure~\ref{fig:Subtree_2_Northern}B2--B3). After the high-level split in Panel~B1, the eastern and southern corridor splits again. In Panel~B2, the the southern settlements becomes an even more poorly served branch, with a mean distance of about 8.0~km, separated from the relatively better-served Yendi corridor with a mean of 4.5~km (see also Appendix Figure~\ref{Appfig_Example_Northern}B2). Panel~B3 further shows the separation of the worse-served the Zabzugu--Tatale district from the better served Yendi urban core(see also Appendix Figure~\ref{Appfig_Example_Northern}B3). Together, these results show that smaller poor-access pockets in Northern are embedded within a wider regional system that is already weakly connected and poorly served.

Taken together,the two cases show that healthcare-access inequality takes different forms across connectivity contexts. In Ashanti, poor access is mainly local: small poor-access pockets appear within an otherwise well-connected region. These pockets are small in territorial extent and usually affect populations at the scale of thousands to tens of thousands. In Northern, poor access are nested across scales: local pockets are embedded within broader regional subsystems that are themselves weakly connected and poorly served. These regional subsystems affect much larger populations, from hundreds of thousands to more than one million people, and extend across broader territorial areas. The contrast shows that poor access is not only a matter of where underserved locations are found, but also of how they are embedded within wider connectivity systems. These different contexts suggest that a single, uniform policy response is unlikely to be effective or targeted enough to address all forms of poor access.

\FloatBarrier
\section{Discussion}\label{sec4}
The results show that healthcare-access inequality in Ghana is not fully captured by asking which locations are poorly served. The street-level analysis answers this first question: it identifies where access is poor and shows that a substantial share of the population remains far from the nearest healthcare facility despite a relatively favourable national mean. However, this point-based view treats each location as an independent observation. As a result, it can miss systemic inequality, where poorly served locations form a poor-access pocket or subsystem separated from the wider and better-served system. The key question therefore shifts from \textit{where are poor-access locations?} to \textit{at what scale does structural healthcare-access inequality emerge?} The hierarchical diagnosis answers this question by identifying the scale at which a spatial concentration of poor access coincides with a connectivity break in the road network. In the results, this appears in different forms: local poor-access pockets, regional poor-access subsystems, and nested pockets across multiple scales.

Methodologically, this study provides an approach for identifying structural healthcare-access inequality through analysing accessibility divergence on the road-network percolation tree. In doing so, it extends the Percolation Divergence Tree framework\citep{Neira2025} from comparing categorical population compositions to comparing empirical distributions of a continuous-valued spatial attribute. This generalisation enables the framework not only to identify multiscale inequalities in healthcare accessibility, but also to be extended to other forms of multiscale spatial inequality and heterogeneity. First, the percolation process groups places according to how they are connected and separated through the road network at different distance thresholds. These connectivity-defined clusters provide endogenous units for comparing healthcare access, rather than imposing exogenous aggregation units, such as administrative boundaries. This reduces dependence on externally defined spatial units and helps address the Modifiable Areal Unit Problem \citep{Openshaw1984}. Second, GJSD is applied along the parent--child relationships of the tree to compare healthcare-access profiles, when smaller clusters merge into larger connected systems. Instead of comparing only mean distance between groups, it compares the full accessibility distributions of child clusters at each merge event \citep{Dagum1997, Attili2020}. This combined approach identifies the meaningful scales where a connectivity break is also associated with clear accessibility divergence: a poor-access pocket or subsystem is structurally separated from a better-served system and has a distinct healthcare-access profile. This matters because many connectivity breaks do not always correspond to meaningful differences in healthcare access.


The higher-level spatial pattern identified by our accessibility analysis also converges with an independent survey-based measure of perceived barriers to healthcare. Ghana's 2022 Demographic and Health Survey reports barriers across several dimensions, including cost, distance to the nearest health facility, permission for treatment, and travelling alone {\citep{GDHS2022}}. Although these self-reported barriers capture a broader dimension of healthcare access than the spatial measure used here, they show a similar regional pattern. In the better-connected and more urbanised southern regions , reported barriers are generally lower. Distance to a health facility is reported as a barrier by 11.7\% of women in Greater Accra and 23.1\% in Ashanti, and by less than 20\% in other surrounding southern regions. This is consistent with our finding that poor spatial access in these relatively advantaged systems is concentrated in small peripheral communities, particularly around metropolitan fringes, with limited population and spatial extent. In contrast, Northern and the surrounding more loosely connected regions report substantially higher barriers on multiple dimensions, particularly for cost and distance, with around 35--41\% reporting distance as a serious problem. This corresponds to the broader and nested regional pattern identified in our analysis, where poor access extends across larger spatial systems rather than isolated local pockets. Importantly, these contrasts are spatially contiguous: lower perceived barriers extend across the southern corridor around Greater Accra and Ashanti, whereas higher barriers extend across Northern, Savannah, North East, and Oti. This correspondence reinforces our central argument that different forms of healthcare-access inequality need to be understood within the broader regional connectivity systems in which they are embedded.


This scale-sensitive diagnosis has direct implications for targeted policy intervention and multi-scale policy coordination. The percolation tree does more than visualise hierarchical connectivity; it indicates the scale at which intervention should operate. Once a poor-access pocket or subsystem is identified, two intervention logics can be considered. The first works \textit{within} the poorly served cluster: where should facilities or services be added or upgraded? Here, the framework defines the target area from road-network connectivity itself, rather than from externally imposed administrative boundaries. The second works \textit{between} clusters: where should connectivity be improved? This is especially important in contexts such as Ghana, where facilities cannot always be distributed, staffed, or maintained in every poorly served location. Connectivity improvement may also address multiple dimensions of access, segregation, and inequality at once, because the same road-network break can affect access to healthcare, education, markets, and other services. The framework therefore helps identify both where local service provision is needed and where connectivity fixes are most urgent.

The two case studies show that different scales of poor access require different forms of intervention. In Ashanti, the problem is mainly local. Poor-access pockets are concentrated around the metropolitan fringe and nearby satellite settlements, while the wider regional system remains relatively well connected and better served. In this context, policy can be more directly targeted at the local scale, either by adding or upgrading facilities within these fringe pockets, or by improving their transport and referral connections to the Kumasi-centred core. Nevertheless, the Northern Region requires a different response. Here, poor access is not only found in local pockets; these pockets are embedded within a wider regional subsystem that is itself weakly connected and poorly served. In such a nested pattern, addressing only the worst local pockets, or only improving short local links, is unlikely to be sustainable solution because the broader regional system also lacks sufficient healthcare resources and connectivity. More effective intervention would require sub-regional and regional coordination, including connectivity improvement between major subsystems, stronger referral routes, and redistribution or coordination of healthcare resources across weakly connected areas. This means that responsibility cannot simply be assigned to local governments with limited capacity to solve a regional structural problem. The policy implication is therefore to match interventions to the scale of poor access: local targeting where poor access is locally concentrated, and multi-level coordination where poor access is regionally embedded.

These findings speak to wider debates on spatial aggregation, settlement hierarchy, and relational inequality. First, by locating the scale at which poor access coincides with road-network separation, the framework shows that aggregation is not only a technical measurement problem but also a question of the ``inequality horizon'' at which inequality emerges and hence intervention should operate \citep{DuranNebreda2025}. Second, the results connect healthcare-access inequality to Africa's bottom-heavy settlement hierarchy, where weakly connected and polycentric regional systems require more targeted and distributed forms of infrastructure and service intervention rather than planning around a single dominant centre \citep{Heinrigs2020}. Finally, the findings support a relational view of spatial inequality: poor access is not only a property of individual locations, but emerges from how places are connected to, or separated from, wider service and settlement systems \citep{Batty2020, RodriguezPose2018, PrietoCuriel2022, Bettencourt2025}.

\subsection*{Limitations}
This study has several limitations. First, population grids were assigned to their nearest road intersections so that population could be incorporated into the road-network structure. This provides a practical way to analyse accessibility within a connected spatial system, but it may underestimate additional barriers faced by households located far from mapped roads or intersections. Second, the road network was reconstructed using OpenStreetMap. Although OSM road data in Africa has uneven completeness \cite{BarringtonLeigh2017}, validation work suggests that it broadly captures Ghana's main road-network structure, with missing-road issues happens in some local grid cells \cite{QualityAssessmentWestAfrica2020}. The resulting network should therefore be interpreted as an approximation of Ghana's road-connectivity backbone rather than a complete representation of all local paths and access routes. Third, healthcare accessibility was measured as network distance to the nearest facility. This provides a consistent national baseline, but it does not account for facility capacity, service quality, opening hours, referral systems, or differences in facility type \citep{Zhou2026}. Communities close to a facility may still face limited effective access if the facility is overloaded, under-resourced, or unable to provide the needed service. Relatedly, and as noted in the Methods, this facility-based measure captures only the fixed-location component of healthcare provision. It does not capture outreach services delivered by Community Health Officers beyond CHPS compounds, which play an important role in rural and poorly accessible areas of Ghana {\citep{Asiedu2023}}. The poor-access pockets identified here should therefore be interpreted as areas with poor access to fixed healthcare facilities, rather than as areas entirely without primary healthcare contact. The poor-access pockets identified in this paper should therefore be interpreted as areas with poor access to fixed facilities, not as areas entirely without primary health-care contact. Fourth, the analysis used walking distance as an equity-oriented baseline and because nationally consistent data on public transport routes, schedules, costs, and reliability are not available. Actual travel behaviour may involve multiple transport modes and may be shaped by road condition, income, gender, household responsibilities, and other socioeconomic constraints. Finally, the percolation procedure depends on threshold step size and stopping criteria. The identified thresholds should therefore be interpreted as structural approximations rather than exact cutoffs, although the GJSD signal helps identify the most meaningful transitions in the hierarchy.

\section{Conclusion}\label{sec5}
This paper developed a connectivity-based multiscale framework for identifying structural healthcare-access inequality and applied it to Ghana. The framework combines road-network percolation with GJSD, following the work introduced by Neira et al \citep{Neira2025}. The percolation process groups places according to how they are connected and separated through the road network at different distance thresholds, while GJSD compares healthcare-access distributions along the parent--child relationships of the percolation tree. This allows the analysis to move beyond asking \textit{Where are poor-access locations?} and instead ask \textit{At what scale does structural healthcare-access inequality emerge?}

The empirical results show three main findings. First, street-level accessibility analysis reveals that Ghana's national average masks substantial inequality. Although the mean distance to the nearest healthcare facility is below the 5~km reference threshold, around one quarter of the population lives more than 5~km from the nearest facility, and nearly one tenth lives more than 10~km away. Second, the road-network percolation tree shows that Ghana's connectivity structure is highly uneven. Some regional systems, such as Ashanti, join the national backbone at relatively low thresholds, while more dispersed northern systems remain separated until higher thresholds. Third, GJSD shows that not every connectivity break produces healthcare-access inequality. The meaningful scales are those where a spatial concentration of poor access coincides with a connectivity break in the road network. These scales appear in different forms: local poor-access pockets, regional poor-access subsystems, and nested pockets embedded within wider poorly served systems.

The case studies further show why this distinction matters. In Ashanti, poor access mainly appears in small peripheral communities around the metropolitan fringe and nearby satellite settlements within an otherwise relatively well-connected system. In the Northern Region, poor access is broader and nested: local pockets are embedded within weakly connected and poorly served regional subsystems. These contrasting patterns are also consistent with regional differences in perceived barriers to healthcare reported in Ghana's 2022 Demographic and Health Survey, reinforcing the importance of interpreting local access inequality within its wider regional context. Different forms of inequality therefore imply different policy responses. Local pockets may be addressed through targeted facility provision, service upgrading, or local connectivity improvements, whereas regional and nested poor-access systems require broader coordination, including connectivity improvement, referral strengthening, and coordination of healthcare resources across weakly connected areas.

Overall, the framework shows that healthcare-access inequality is not only a matter of distance from facilities, but also of how poorly served places are connected to, or separated from, wider service systems. This scale-sensitive perspective can help identify both where inequalities are concentrated and the spatial level at which intervention may need to operate. The approach can also be extended to other infrastructure and service-accessibility domains, including education, markets, water, and emergency services. Future work could incorporate facility capacity, service quality, multimodal travel behaviour, and mobile or outreach-based healthcare provision to provide a fuller account of effective access across connected spatial systems.

More broadly, the approach can contribute to the study of spatial inequality beyond accessibility. Because GJSD can be used to compare the distributions of different spatial variables on the connectivity hierarchy, the framework can help identify the scales at which structural inequality emerges. This opens a way to examine how connectivity breaks relate to nested forms of deprivation across multiple dimensions and scales, including health, safety, income, and economic opportunity. By identifying where disadvantaged communities remain locally or regionally disconnected from wider service and opportunity systems, such approaches can support more spatially targeted efforts to reduce inequalities in access and advance universal health coverage.


\backmatter

\bmhead{Supplementary information}


Additional figures, tables, and analyses that support the findings of this study are provided in the Supplementary Information. These include extended methodological details and supplementary figures.

\bmhead{Acknowledgements}

\section*{Declarations}



This project is conducted in collaboration with the Malaria Atlas Project at the Kids Research Institute, Australia.

\counterwithin{figure}{section}
\counterwithin{table}{section}
\clearpage

\begin{appendices}





\section{Supplementary Figures and Analyses}\label{secA1}

This appendix presents supplementary material that provides additional context and supporting evidence for the main analysis.

\begin{figure}[H]
  \centering
  \includegraphics[width=0.8 \textwidth]{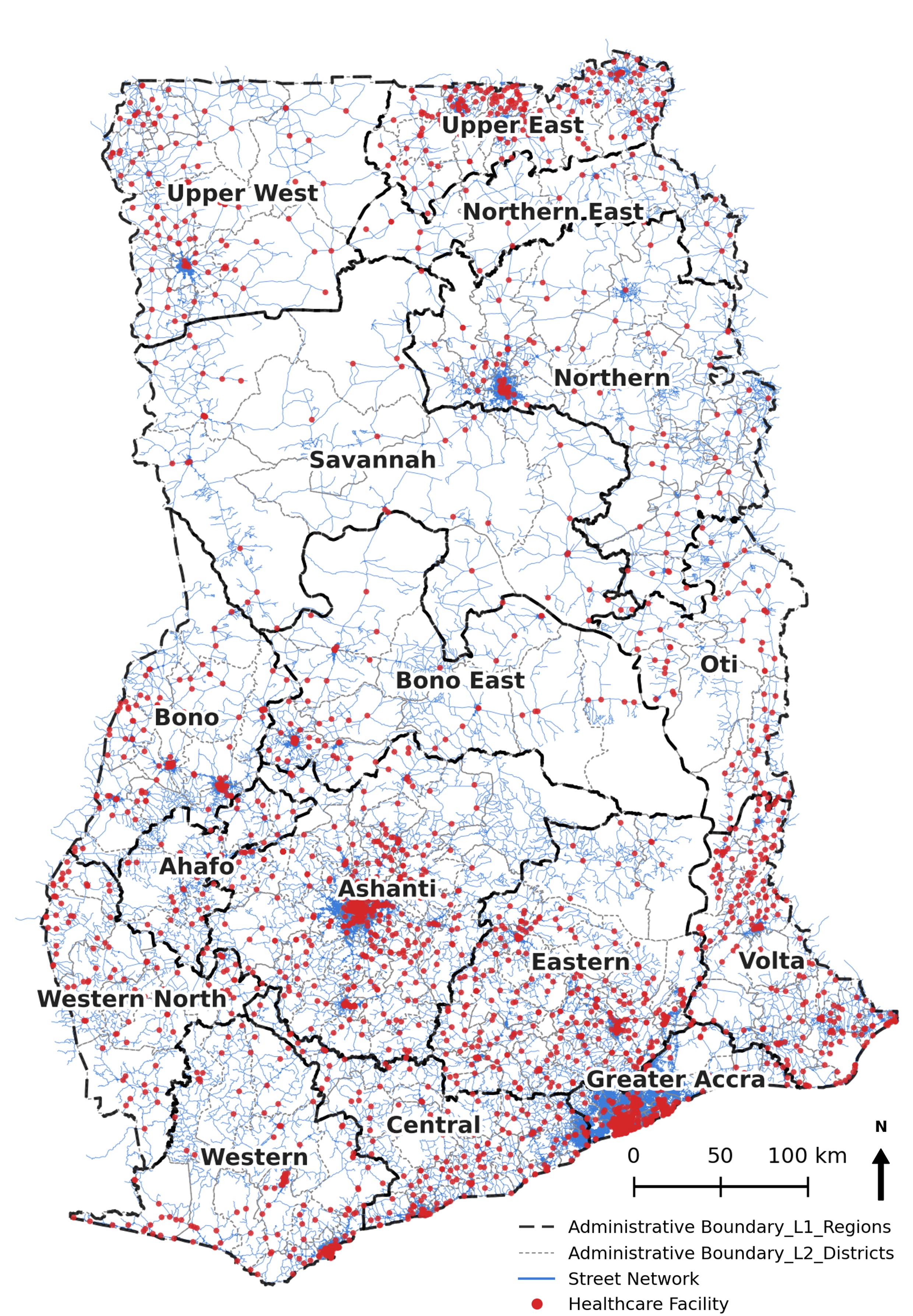}
  \caption{Ghana Administrative structure/Healthcare services.}
  \label{Appfig_Healtcare_Facilities}
\end{figure}

\newpage

\begin{figure}[H]
  \centering
  \includegraphics[width=\textwidth]{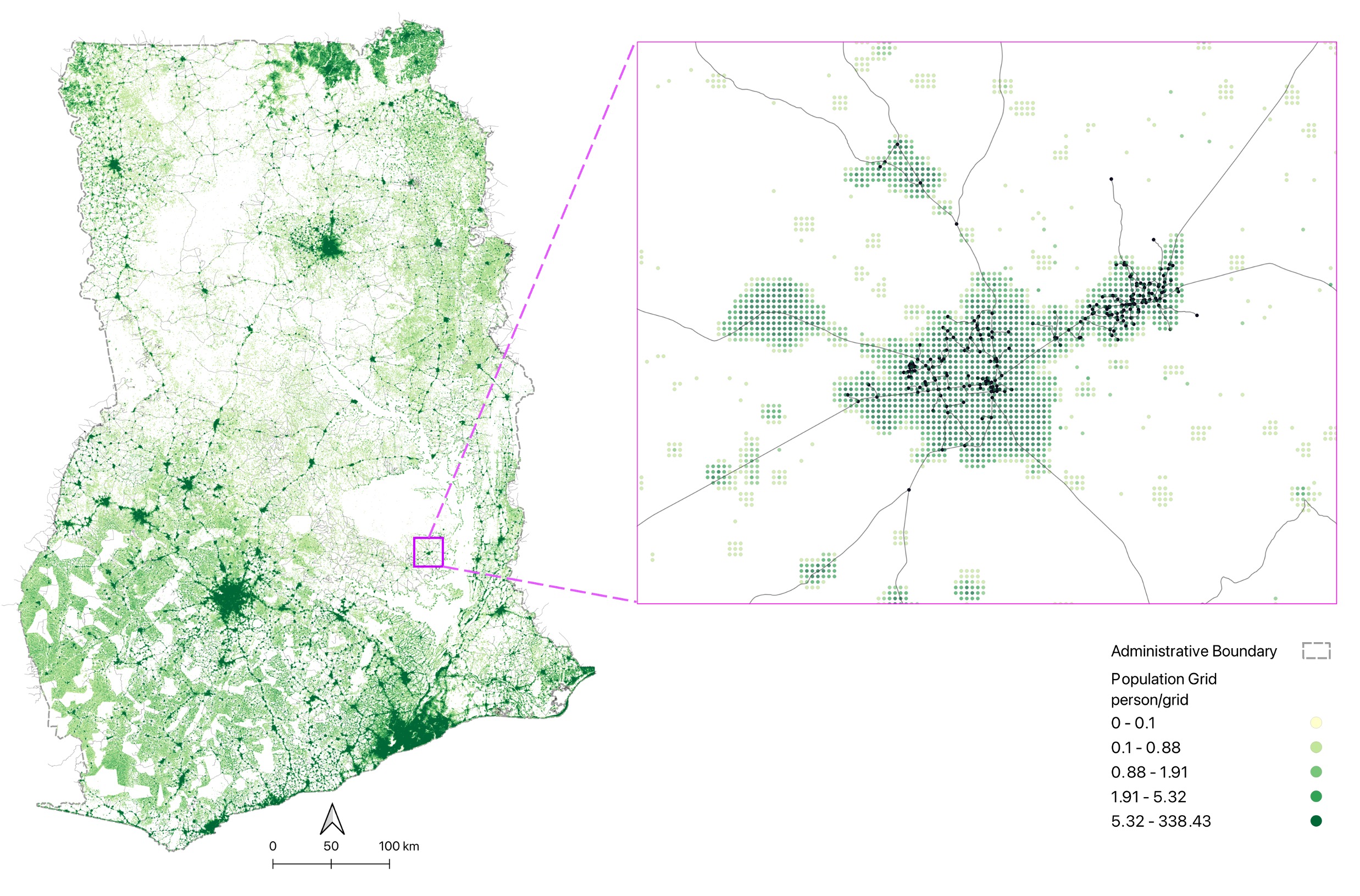}
  \caption{ Worldpop Map. WorldPop 2024 population grid used for assigning population weights to street intersections.}
  \label{Appfig:worldpop_map}
\end{figure}

\newpage

\begin{figure}[H]
  \centering
  \includegraphics[width=0.99\textwidth]{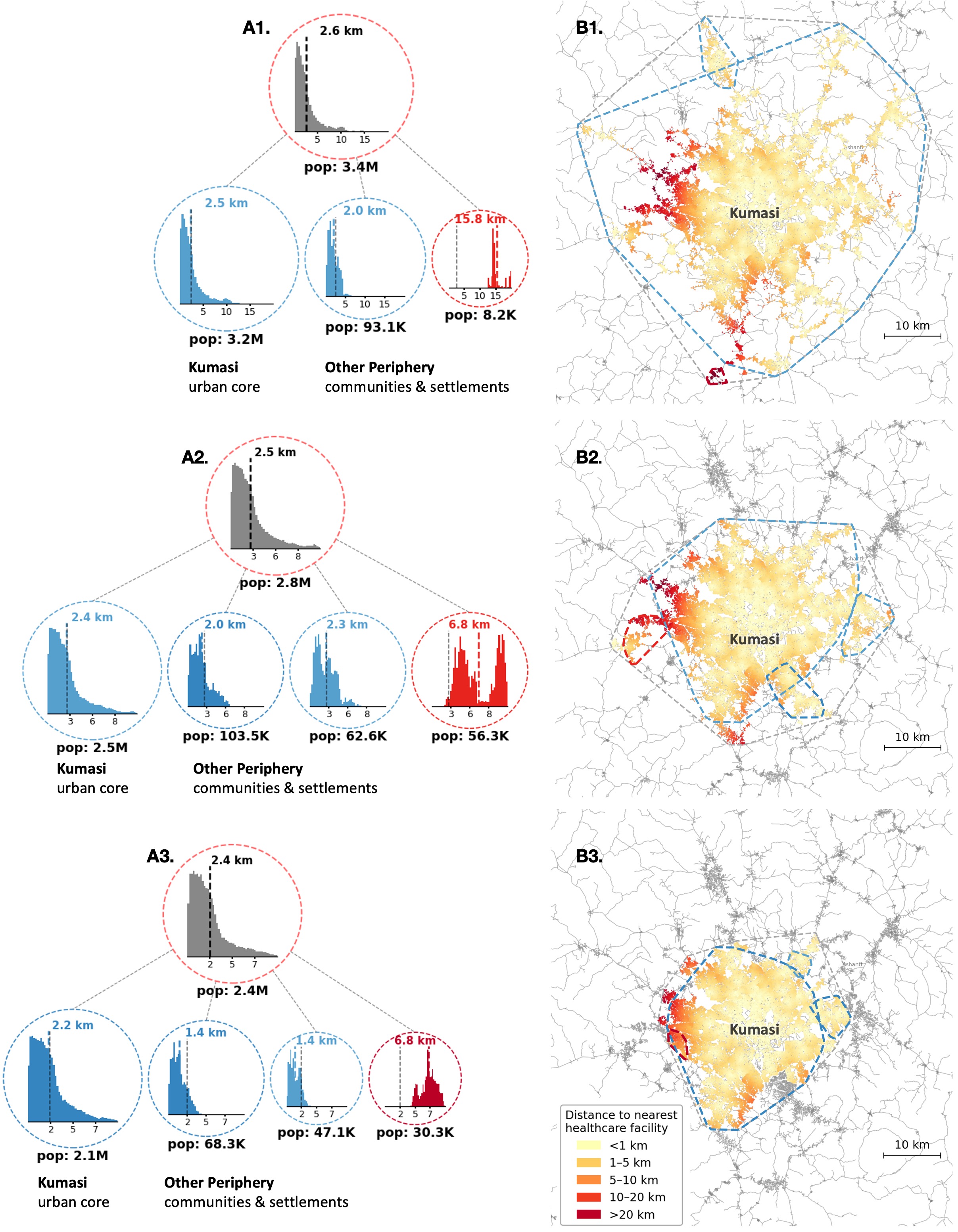}
  \caption{\textbf{Geographic context of flagged scales in Ashanti.}
  \textbf{Panels A1--A3} show the healthcare-access distributions of three selected flagged merge events in the Ashanti subtree, same as Figure~\ref{fig:Subtree_1_Ashanti}B1--B3.
  \textbf{Panels B1--B3} show the corresponding geographic extent of these merge events. Dashed convex hulls delineate the parent cluster in grey and the child clusters in colours, with child-cluster colours indicating mean accessibility. Road intersections are coloured its healthcare accessibility. Across the three examples, the largest child cluster corresponds to the Kumasi urban core, while the smaller worse-served child clusters represent peripheral communities and settlements that merge with the core at different distance thresholds.}
  \label{Appfig_Example_Ashanti}
\end{figure}
\newpage

\begin{figure}[H]
  \centering
  \includegraphics[width= 0.99\textwidth]{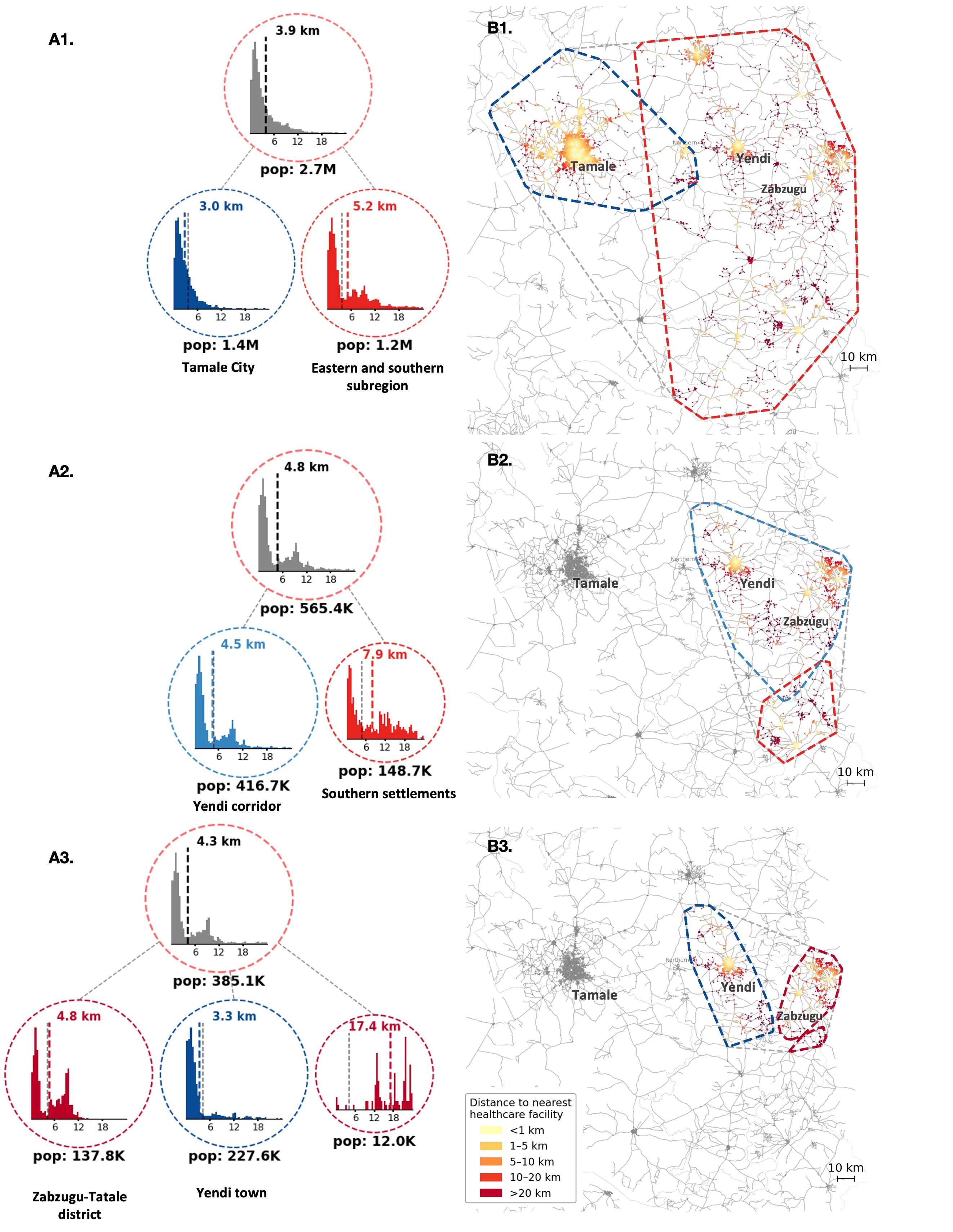}
  \caption{\textbf{Geographic context of flagged scales in the Northern Region.}
  \textbf{Panels A1--A3} show the healthcare-access distributions of three selected flagged merge events in the Northern Region subtree, same as Figure~\ref{fig:Subtree_2_Northern}B1--B3.
  \textbf{Panels B1--B3} show the corresponding geographic extent of these merge events. Dashed convex hulls delineate the parent cluster in grey and the child clusters in colours, with child-cluster colours indicating mean accessibility. Road intersections are coloured its healthcare accessibility. Across the three examples, the maps trace a nested pattern of poor-access inequality: the high-level split separates the worse-served eastern and southern subregion from the better-served Tamale-centred subsystem, while lower-level splits further separate the better-served Yendi urban core from poor-access pockets around the southern settlements and the Zabzugu--Tatale district.}
  \label{Appfig_Example_Northern}
\end{figure}

\end{appendices}

\newpage
\bibliography{ref}

\begin{thebibliography}{10}
\expandafter\ifx\csname url\endcsname\relax
  \def\url#1{\burl{#1}}\fi
\expandafter\ifx\csname urlprefix\endcsname\relax\def\urlprefix{URL }\fi
\providecommand{\bibinfo}[2]{#2}
\providecommand{\eprint}[2][]{\url{#2}}
\providecommand{\doi}[1]{\url{https://doi.org/#1}}
\bibcommenthead

\bibitem{WHOGHOSDG38}
\bibinfo{author}{{World Health Organization}}.
\newblock \bibinfo{title}{{SDG} target 3.8: Achieve universal health coverage ({UHC}), including financial risk protection}.
\newblock \bibinfo{howpublished}{WHO Global Health Observatory} (\bibinfo{year}{2024}).
\newblock \urlprefix\url{https://www.who.int/data/gho/data/themes/topics/indicator-groups/indicator-group-details/GHO/sdg-target-3.8-achieve-universal-health-coverage-(uhc)-including-financial-risk-protection}.

\bibitem{Asamoah2026}
\bibinfo{author}{Asamoah, S.} \emph{et~al.}
\newblock \bibinfo{title}{Universal health coverage and health equity in {West Africa}: Tracking progress toward {SDG} 3.8 in {Ghana} and {Liberia}}.
\newblock \emph{\bibinfo{journal}{Public Health Challenges}} \textbf{\bibinfo{volume}{5}}, \bibinfo{pages}{e70280} (\bibinfo{year}{2026}).

\bibitem{Falchetta2020}
\bibinfo{author}{Falchetta, G.} \emph{et~al.}
\newblock \bibinfo{title}{Planning universal accessibility to public health care in sub-{Saharan Africa}}.
\newblock \emph{\bibinfo{journal}{Proceedings of the National Academy of Sciences}} \textbf{\bibinfo{volume}{117}}, \bibinfo{pages}{31761--31769} (\bibinfo{year}{2020}).

\bibitem{Bhatt2015}
\bibinfo{author}{Bhatt, S.} \emph{et~al.}
\newblock \bibinfo{title}{The effect of malaria control on {Plasmodium falciparum} in {Africa} between 2000 and 2015}.
\newblock \emph{\bibinfo{journal}{Nature}} \textbf{\bibinfo{volume}{526}}, \bibinfo{pages}{207--211} (\bibinfo{year}{2015}).

\bibitem{Deaton2001}
\bibinfo{author}{Deaton, A.}
\newblock \bibinfo{title}{Relative deprivation, inequality, and mortality}.
\newblock \bibinfo{type}{Working Paper} \bibinfo{number}{8099}, \bibinfo{institution}{National Bureau of Economic Research} (\bibinfo{year}{2001}).

\bibitem{Chetty2026}
\bibinfo{author}{Chetty, R.}, \bibinfo{author}{Friedman, J.~N.}, \bibinfo{author}{Hendren, N.}, \bibinfo{author}{Jones, M.~R.} \& \bibinfo{author}{Porter, S.~R.}
\newblock \bibinfo{title}{The opportunity atlas: Mapping the childhood roots of social mobility}.
\newblock \emph{\bibinfo{journal}{American Economic Review}} \textbf{\bibinfo{volume}{116}}, \bibinfo{pages}{1--51} (\bibinfo{year}{2026}).

\bibitem{Sarkar2024}
\bibinfo{author}{Sarkar, S.}, \bibinfo{author}{Cottineau-Mugadza, C.} \& \bibinfo{author}{Wolf, L.~J.}
\newblock \bibinfo{title}{Spatial inequalities and cities: A review}.
\newblock \emph{\bibinfo{journal}{Environment and Planning B: Urban Analytics and City Science}} \textbf{\bibinfo{volume}{51}}, \bibinfo{pages}{1391--1407} (\bibinfo{year}{2024}).

\bibitem{Chetty2022SocialCapital}
\bibinfo{author}{Chetty, R.} \emph{et~al.}
\newblock \bibinfo{title}{Social capital i: Measurement and associations with economic mobility}.
\newblock \emph{\bibinfo{journal}{Nature}} \textbf{\bibinfo{volume}{608}}, \bibinfo{pages}{108--121} (\bibinfo{year}{2022}).

\bibitem{Arcaute2016}
\bibinfo{author}{Arcaute, E.} \emph{et~al.}
\newblock \bibinfo{title}{Cities and regions in {Britain} through hierarchical percolation}.
\newblock \emph{\bibinfo{journal}{Royal Society Open Science}} \textbf{\bibinfo{volume}{3}}, \bibinfo{pages}{150691} (\bibinfo{year}{2016}).

\bibitem{RodriguezPose2018}
\bibinfo{author}{Rodr{\'i}guez-Pose, A.}
\newblock \bibinfo{title}{The revenge of the places that don't matter (and what to do about it)}.
\newblock \emph{\bibinfo{journal}{Cambridge Journal of Regions, Economy and Society}} \textbf{\bibinfo{volume}{11}}, \bibinfo{pages}{189--209} (\bibinfo{year}{2018}).

\bibitem{Wu2025}
\bibinfo{author}{Wu, S.} \emph{et~al.}
\newblock \bibinfo{title}{Measuring global human accessibility to essential daily necessities and services}.
\newblock \emph{\bibinfo{journal}{Nature Communications}} \textbf{\bibinfo{volume}{16}} (\bibinfo{year}{2025}).

\bibitem{Weiss2018}
\bibinfo{author}{Weiss, D.~J.} \emph{et~al.}
\newblock \bibinfo{title}{A global map of travel time to cities to assess inequalities in accessibility in 2015}.
\newblock \emph{\bibinfo{journal}{Nature}} \textbf{\bibinfo{volume}{553}}, \bibinfo{pages}{333--336} (\bibinfo{year}{2018}).

\bibitem{Pandey2025}
\bibinfo{author}{Pandey, B.}, \bibinfo{author}{Brelsford, C.} \& \bibinfo{author}{Seto, K.~C.}
\newblock \bibinfo{title}{Rising infrastructure inequalities accompany urbanization and economic development}.
\newblock \emph{\bibinfo{journal}{Nature Communications}} \textbf{\bibinfo{volume}{16}} (\bibinfo{year}{2025}).

\bibitem{Tu2025}
\bibinfo{author}{Tu, Y.} \emph{et~al.}
\newblock \bibinfo{title}{Inequality in infrastructure access and its association with health disparities}.
\newblock \emph{\bibinfo{journal}{Nature Human Behaviour}} \textbf{\bibinfo{volume}{9}}, \bibinfo{pages}{1669--1682} (\bibinfo{year}{2025}).

\bibitem{Bruno2024}
\bibinfo{author}{Bruno, M.} \emph{et~al.}
\newblock \bibinfo{title}{A universal framework for inclusive 15-minute cities}.
\newblock \emph{\bibinfo{journal}{Nature Cities}} \textbf{\bibinfo{volume}{1}}, \bibinfo{pages}{633--641} (\bibinfo{year}{2024}).

\bibitem{Zhou2026}
\bibinfo{author}{Zhou, S.} \emph{et~al.}
\newblock \bibinfo{title}{The moving target of urban equity: Spatiotemporal demand and double disadvantage in {Hefei}, {China}}.
\newblock \emph{\bibinfo{journal}{arXiv preprint arXiv:2606.20132}}  (\bibinfo{year}{2026}).

\bibitem{LuoWang2003}
\bibinfo{author}{Luo, W.} \& \bibinfo{author}{Wang, F.}
\newblock \bibinfo{title}{Measures of spatial accessibility to health care in a {GIS} environment}.
\newblock \emph{\bibinfo{journal}{Environment and Planning B: Planning and Design}} \textbf{\bibinfo{volume}{30}}, \bibinfo{pages}{865--884} (\bibinfo{year}{2003}).

\bibitem{Wang2012}
\bibinfo{author}{Wang, F.}
\newblock \bibinfo{title}{Measurement, optimization, and impact of health care accessibility: A methodological review}.
\newblock \emph{\bibinfo{journal}{Annals of the Association of American Geographers}} \textbf{\bibinfo{volume}{102}}, \bibinfo{pages}{1104--1112} (\bibinfo{year}{2012}).

\bibitem{Tao2020}
\bibinfo{author}{Tao, Z.}, \bibinfo{author}{Cheng, Y.} \& \bibinfo{author}{Liu, J.}
\newblock \bibinfo{title}{Hierarchical two-step floating catchment area ({2SFCA}) method: measuring the spatial accessibility to hierarchical healthcare facilities in {Shenzhen, China}}.
\newblock \emph{\bibinfo{journal}{International Journal for Equity in Health}} \textbf{\bibinfo{volume}{19}}, \bibinfo{pages}{164} (\bibinfo{year}{2020}).

\bibitem{Openshaw1984}
\bibinfo{author}{Openshaw, S.}
\newblock \bibinfo{title}{Ecological fallacies and the analysis of areal census data}.
\newblock \emph{\bibinfo{journal}{Environment and Planning A: Economy and Space}} \textbf{\bibinfo{volume}{16}}, \bibinfo{pages}{17--31} (\bibinfo{year}{1984}).

\bibitem{Attili2020}
\bibinfo{author}{Attili, F.}
\newblock \bibinfo{title}{Within-between decomposition of the {Gini} index: A novel proposal}.
\newblock \emph{\bibinfo{journal}{Quaderni --- Working Paper DSE}}  (\bibinfo{year}{2020}).

\bibitem{Marin2024}
\bibinfo{author}{Marin, V.} \emph{et~al.}
\newblock \bibinfo{title}{The scalar mismatch of regional governance}.
\newblock \emph{\bibinfo{journal}{Environment and Planning B: Urban Analytics and City Science}} \textbf{\bibinfo{volume}{51}}, \bibinfo{pages}{2126--2145} (\bibinfo{year}{2024}).

\bibitem{Jia2017}
\bibinfo{author}{Jia, P.}, \bibinfo{author}{Wang, F.} \& \bibinfo{author}{Xierali, I.~M.}
\newblock \bibinfo{title}{Delineating hierarchical hospital service areas in {Florida}}.
\newblock \emph{\bibinfo{journal}{Geographical Review}} \textbf{\bibinfo{volume}{107}}, \bibinfo{pages}{608--623} (\bibinfo{year}{2017}).

\bibitem{Dagum1997}
\bibinfo{author}{Dagum, C.}
\newblock \bibinfo{title}{A new approach to the decomposition of the {Gini} income inequality ratio}.
\newblock \emph{\bibinfo{journal}{Empirical Economics}} \textbf{\bibinfo{volume}{22}}, \bibinfo{pages}{515--531} (\bibinfo{year}{1997}).

\bibitem{Silber1989}
\bibinfo{author}{Silber, J.}
\newblock \bibinfo{title}{Factor components, population subgroups and the computation of the {Gini} index of inequality}.
\newblock \emph{\bibinfo{journal}{The Review of Economics and Statistics}} \textbf{\bibinfo{volume}{71}}, \bibinfo{pages}{107--115} (\bibinfo{year}{1989}).

\bibitem{ReySmith2013}
\bibinfo{author}{Rey, S.~J.} \& \bibinfo{author}{Smith, R.~J.}
\newblock \bibinfo{title}{A spatial decomposition of the {Gini} coefficient}.
\newblock \emph{\bibinfo{journal}{Letters in Spatial and Resource Sciences}} \textbf{\bibinfo{volume}{6}}, \bibinfo{pages}{55--70} (\bibinfo{year}{2013}).

\bibitem{Pradhan2003}
\bibinfo{author}{Pradhan, M.}, \bibinfo{author}{Sahn, D.~E.} \& \bibinfo{author}{Younger, S.~D.}
\newblock \bibinfo{title}{Decomposing world health inequality}.
\newblock \emph{\bibinfo{journal}{Journal of Health Economics}} \textbf{\bibinfo{volume}{22}}, \bibinfo{pages}{271--293} (\bibinfo{year}{2003}).

\bibitem{KaplanOmer2022}
\bibinfo{author}{Kaplan, N.} \& \bibinfo{author}{Omer, I.}
\newblock \bibinfo{title}{Multiscale accessibility---{A} new perspective of space structuration}.
\newblock \emph{\bibinfo{journal}{Sustainability}} \textbf{\bibinfo{volume}{14}}, \bibinfo{pages}{5119} (\bibinfo{year}{2022}).

\bibitem{DuranNebreda2025}
\bibinfo{author}{Duran-Nebreda, S.}, \bibinfo{author}{Vidiella, B.}, \bibinfo{author}{Bentley, R.~A.} \& \bibinfo{author}{Valverde, S.}
\newblock \bibinfo{title}{Fractal clusters and urban scaling shape spatial inequality in {U.S.} patenting}.
\newblock \emph{\bibinfo{journal}{npj Complexity}} \textbf{\bibinfo{volume}{2}} (\bibinfo{year}{2025}).

\bibitem{Heinrigs2020}
\bibinfo{author}{Heinrigs, P.}
\newblock \bibinfo{title}{Africapolis: understanding the dynamics of urbanization in {Africa}}.
\newblock \emph{\bibinfo{journal}{Field Actions Science Reports}} \textbf{\bibinfo{volume}{22}}, \bibinfo{pages}{18--23} (\bibinfo{year}{2020}).

\bibitem{Xu2025}
\bibinfo{author}{Xu, G.} \emph{et~al.}
\newblock \bibinfo{title}{Underlying rules of evolutionary urban systems in {Africa}}.
\newblock \emph{\bibinfo{journal}{Nature Cities}} \textbf{\bibinfo{volume}{2}}, \bibinfo{pages}{327--335} (\bibinfo{year}{2025}).

\bibitem{Bettencourt2025}
\bibinfo{author}{Bettencourt, L. M.~A.} \& \bibinfo{author}{Marchio, N.}
\newblock \bibinfo{title}{Infrastructure deficits and informal settlements in sub-{Saharan Africa}}.
\newblock \emph{\bibinfo{journal}{Nature}} \textbf{\bibinfo{volume}{645}}, \bibinfo{pages}{399--406} (\bibinfo{year}{2025}).

\bibitem{PrietoCuriel2022}
\bibinfo{author}{Prieto-Curiel, R.}, \bibinfo{author}{Schumann, A.}, \bibinfo{author}{Heo, I.} \& \bibinfo{author}{Heinrigs, P.}
\newblock \bibinfo{title}{Detecting cities with high intermediacy in the {African} urban network}.
\newblock \emph{\bibinfo{journal}{Computers, Environment and Urban Systems}} \textbf{\bibinfo{volume}{98}}, \bibinfo{pages}{101869} (\bibinfo{year}{2022}).

\bibitem{Neira2025}
\bibinfo{author}{Neira, M.}, \bibinfo{author}{Marin, V.} \& \bibinfo{author}{Arcaute, E.}
\newblock \bibinfo{title}{Multiscalarity in socio-spatial segregation: An information-theoretic framework}.
\newblock \emph{\bibinfo{journal}{arXiv preprint arXiv:2505.14937}}  (\bibinfo{year}{2025}).

\bibitem{Rozenfeld2011}
\bibinfo{author}{Rozenfeld, H.~D.}, \bibinfo{author}{Rybski, D.}, \bibinfo{author}{Gabaix, X.} \& \bibinfo{author}{Makse, H.~A.}
\newblock \bibinfo{title}{The area and population of cities: New insights from a different perspective on cities}.
\newblock \emph{\bibinfo{journal}{American Economic Review}} \textbf{\bibinfo{volume}{101}}, \bibinfo{pages}{2205--2225} (\bibinfo{year}{2011}).

\bibitem{Murcio2015}
\bibinfo{author}{Murcio, R.}, \bibinfo{author}{Morphet, R.}, \bibinfo{author}{Gershenson, C.} \& \bibinfo{author}{Batty, M.}
\newblock \bibinfo{title}{Urban transfer entropy across scales}.
\newblock \emph{\bibinfo{journal}{PLOS ONE}} \textbf{\bibinfo{volume}{10}}, \bibinfo{pages}{e0133780} (\bibinfo{year}{2015}).

\bibitem{Murcio2025}
\bibinfo{author}{Murcio, R.} \& \bibinfo{author}{Soundararaj, B.}
\newblock \bibinfo{title}{Trends in urban flows: A transfer entropy approach}.
\newblock \emph{\bibinfo{journal}{arXiv preprint arXiv:2501.06316}}  (\bibinfo{year}{2025}).

\bibitem{Asiedu2023}
\bibinfo{author}{Asiedu, A.} \emph{et~al.}
\newblock \bibinfo{title}{Improving malaria case management and referral relationships at the primary care level in {Ghana}: Evaluation of a quality assurance internship}.
\newblock \emph{\bibinfo{journal}{Global Health: Science and Practice}} \textbf{\bibinfo{volume}{11}}, \bibinfo{pages}{e2300050} (\bibinfo{year}{2023}).

\bibitem{Tatem2017WorldPop}
\bibinfo{author}{Tatem, A.~J.}
\newblock \bibinfo{title}{{WorldPop}, open data for spatial demography}.
\newblock \emph{\bibinfo{journal}{Scientific Data}} \textbf{\bibinfo{volume}{4}}, \bibinfo{pages}{170004} (\bibinfo{year}{2017}).

\bibitem{WorldPopR2024B}
\bibinfo{author}{Bondarenko, M.} \emph{et~al.}
\newblock \bibinfo{title}{Constrained estimates of 2015--2030 total number of people per grid square at a resolution of 3 arc-seconds (approximately 100~m at the equator), {R2024B} version v1}.
\newblock \bibinfo{howpublished}{WorldPop, School of Geography and Environmental Science, University of Southampton} (\bibinfo{year}{2024}).
\newblock \bibinfo{note}{Global Demographic Data Project, funded by the Bill \& Melinda Gates Foundation (INV-045237). Ghana 2023 constrained estimate accessed via \url{https://data.worldpop.org/GIS/Population/Global_2015_2030/R2024B/2023/GHA/v1/100m/constrained/}}.

\bibitem{WHO2019PHCUHC}
\bibinfo{author}{{World Health Organization}}.
\newblock \bibinfo{title}{Primary health care on the road to universal health coverage: 2019 monitoring report} (\bibinfo{year}{2019}).
\newblock \urlprefix\url{https://www.who.int/publications/i/item/9789240029040}.

\bibitem{GDHS2022}
\bibinfo{author}{{Ghana Statistical Service (GSS)}}, \bibinfo{author}{{Ghana Health Service (GHS)}} \& \bibinfo{author}{{ICF}}.
\newblock \bibinfo{title}{Ghana demographic and health survey 2022} (\bibinfo{year}{2023}).
\newblock \urlprefix\url{https://dhsprogram.com/pubs/pdf/FR387/FR387.pdf}.

\bibitem{Batty2020}
\bibinfo{author}{Batty, M.}
\newblock \bibinfo{title}{On scale and size}.
\newblock \emph{\bibinfo{journal}{Environment and Planning B: Urban Analytics and City Science}} \textbf{\bibinfo{volume}{47}}, \bibinfo{pages}{359--362} (\bibinfo{year}{2020}).

\bibitem{BarringtonLeigh2017}
\bibinfo{author}{Barrington-Leigh, C.} \& \bibinfo{author}{Millard-Ball, A.}
\newblock \bibinfo{title}{The world’s user-generated road map is more than 80\% complete}.
\newblock \emph{\bibinfo{journal}{PLOS ONE}} \textbf{\bibinfo{volume}{12}}, \bibinfo{pages}{e0180698} (\bibinfo{year}{2017}).

\bibitem{QualityAssessmentWestAfrica2020}
\bibinfo{author}{Cîrnugea, A.}, \bibinfo{author}{Herfort, B.} \& \bibinfo{author}{Zipf, A.}
\newblock \bibinfo{title}{Quality assessment of crowd-sourced data: {OpenStreetMap} roads validation in the developing countries of {West Africa}}.
\newblock \emph{\bibinfo{journal}{Geo-spatial Information Science}} \textbf{\bibinfo{volume}{24}}, \bibinfo{pages}{349--367} (\bibinfo{year}{2021}).

\end{thebibliography}

\end{document}